\documentclass[9pt]{article}

\usepackage[fontsize=9pt]{scrextend}
\usepackage{url}
\usepackage{soul}

\usepackage{empheq}
\usepackage{mathtools}
\usepackage{multirow}
\usepackage{adjustbox}

\usepackage{booktabs} 
\usepackage{xfrac}
\usepackage{ragged2e}
\justifying

\usepackage{multicol}
\usepackage{amssymb}

\NewDocumentCommand{\overarrow}{O{=} O{\uparrow} m}{%
  \overset{\makebox[0pt]{\begin{tabular}{@{}c@{}}#3\\[0pt]\ensuremath{#2}\end{tabular}}}{#1}
}
\NewDocumentCommand{\underarrow}{O{=} O{\downarrow} m}{%
  \underset{\makebox[0pt]{\begin{tabular}{@{}c@{}}\ensuremath{#2}\\[0pt]#3\end{tabular}}}{#1}
}

\usepackage{mathrsfs}
\usepackage[tableposition=top]{caption}
\usepackage[flushleft]{threeparttable}
\usepackage{floatrow}
\usepackage{algorithm}      
\usepackage{algpseudocode}  
\usepackage{xcolor}
\usepackage{amsmath}
\usepackage[most]{tcolorbox}

\usepackage{tikz}
\usetikzlibrary{positioning,fit,calc,arrows.meta}

\usepackage{geometry}
\usepackage{enumitem}
\newcommand{\psup}{(\mkern-1.5mu \ell\mkern-2mu)} 
\usepackage{titlesec}
\titlespacing{\paragraph}{%
  0pt}{
  0.5em}{
  1em}
\date{}

\usepackage{authblk}
\usepackage{hyphenat}

\usepackage[square,numbers,sort&compress]{natbib}
\usepackage[T1]{fontenc}
\usepackage[utf8]{inputenc}
\usepackage{newtxtext}          
\usepackage{newtxmath}

\usepackage{bm,accents}
\usepackage{pifont}

\usepackage{graphicx}

\DeclareGraphicsExtensions{.pdf}
\usepackage{amsmath}

\title{\textbf{Learning inelastic constitutive models from stress--strain data \\
under hard thermodynamic constraints}}

{
\author{Filippo Masi}
{\affil{\itshape Univ. Grenoble Alpes, Inria, CNRS, Grenoble INP, Institute of Engineering, LJK, 38000 Grenoble, France.
}}
}
\begin{document}
\maketitle
\begingroup
\renewcommand{\thefootnote}{}
\footnotetext{\\ \vspace{-15pt}

\noindent F.~Masi, ``Learning inelastic constitutive models from stress--strain data under hard thermodynamic constraints,'' \emph{Computer Methods in Applied Mechanics and Engineering}, 461 (2026) 119260. doi: \url{https://doi.org/10.1016/j.cma.2026.119260}.}
\addtocounter{footnote}{-1}
\endgroup

\subsection*{Abstract}

Machine learning approaches informed by physics have offered new insights into the discovery of constitutive models from data, helping overcome some limitations of traditional constitutive modelling while reducing the cost of otherwise computationally intensive simulations. Yet, many existing methods either enforce only part of the relevant physical and thermodynamic structure, or achieve thermodynamic consistency within specific constitutive classes, leaving open questions about their applicability across a broad range of material behaviours and their ability to generalise to unseen loading paths when limited data are available.

This work establishes a thermodynamics-constrained learning framework for inelastic constitutive models from macroscopic stress--strain data. The framework is based on non-equilibrium thermodynamics and parameterises both the free energy and the generalised transport operator governing the evolution of the material state, while enforcing material objectivity, energy balance, and non-negative dissipation as hard, scalable constraints. Analytical benchmarks involving simple stress--strain loading paths demonstrate that the method learns thermodynamically consistent and robust constitutive models for a range of inelastic materials of increasing complexity. At inference, the resulting models generalise to more demanding, unobserved paths and identify interpretable internal variables that capture path-dependent behaviours. The framework is then applied to granular media, prototypical heterogeneous and history-dependent materials. Trained on numerically simulated experiments based on the discrete element method, the method identifies admissible constitutive equations and predicts the response under cyclic loading, including the emergence of hysteresis absent from the training data, relying solely on macroscopic stress--strain histories.

\paragraph*{Keywords} Machine learning; Data-driven constitutive modelling; Hard-constrained learning; Non-equilibrium thermodynamics; Transport equations; Granular materials.

\subsection*{Highlights}
\begin{itemize}
\itemsep-0.3em 
    \item Non-equilibrium thermodynamics imposed as hard constraints.
    \item Inelastic constitutive models identified from simple stress--strain data.
    \item Learned models generalise to demanding loading paths beyond the training set.
    \item Interpretable internal variables are identified.
    \item Application to history-dependent granular media with high-fidelity, grain-scale simulations.
\end{itemize}

\section{Introduction}
\label{sec:introduction}

A central objective in solid mechanics and materials science is to predict and understand the response of materials to external stimuli. In continuum mechanics, this predictive capability hinges on the constitutive closure, which complements kinematics and balance equations (e.g.\ momentum balance) and links observable measures of deformation to stresses and to the evolution of the material state. Beyond accuracy, constitutive models must satisfy structural requirements dictated by physics, including frame indifference and consistency with the laws of thermodynamics. Yet, for materials displaying path-dependent behaviours and intrinsic microstructural effects, traditional heuristic constitutive models are often hindered by phenomenological choices of the state variables and the evolution equations, typically refined through human trial-and-error adjustments. Multiscale modelling overcomes this issue by replacing heuristic idealisations with implicit constitutive closures obtained from direct numerical simulations of nested boundary-value problems posed on representative elementary volumes (auxiliary problems) and the subsequent upscaling of the effective behaviour \citep{geers2010multi}. However, the high computational cost of these methods, due to the repeated, time-consuming solution of the auxiliary problem, limits their use in real-scale applications.

To address this gap, machine learning and data-driven methods have been increasingly explored to discover constitutive equations and build high-fidelity surrogate models directly from available experimental measurements and numerical simulations (for an extensive review, see \citep{montans2019data,fuhg2025review}). A central development is the integration of first principles directly into learning and identification procedures to improve robustness and generalisation while promoting physically consistent predictions (among others, see \citep{karniadakis2021physics,karapiperis2021data,masi2021thermodynamics,hernandez2021deep,cueto2023thermodynamics,klein2022polyconvex,masi2022multiscale,flaschel2023automated,linden2023neural,as2023mechanics,linka2023new,rosenkranz2024viscoelasticty,zlatic2024data,qiu2025bridging,jailin2025material,flaschel2025convex}). A key design choice is how physical principles are imposed: through soft constraints (penalisation in the loss) or hard constraints (admissibility by construction). Soft constraints are straightforward to integrate into standard architectures and optimisation algorithms \citep{karniadakis2021physics}, but do not guarantee the fulfilment of the physical principles at inference, particularly under extrapolation outside the training distribution. This can lead to ``silent'' failures, i.e., plausible yet unphysical predictions without an explicit indicator of inconsistency. Hard constraints, in contrast, enforce the prescribed principles exactly during training and at inference by restricting the hypothesis class a priori, which is found to strengthen generalisation capabilities while providing guarantees of consistency. The trade-off is reduced flexibility: hard constraints often require tailored parameterisations and may limit predictive performance if the imposed structure is overly restrictive or does not align with the observed material behaviour. In practice, both formulations have proved valuable, and hybrid strategies are now common for their flexibility.

In elasticity, several successful approaches learn free-energy potentials using objective inputs, polyconvex parametrisations, and Sobolev training \citep{czarnecki2017sobolev} to predict the material stress by automatic differentiation, while explicitly enforcing material objectivity and energy balance (cf. \citep{klein2022polyconvex,chen2022polyconvex,linka2023new,linden2023neural}). For inelastic and complex materials, an increasing number of methods leverage plasticity-inspired formulations and the internal variable approach to account for the path-dependent and multiscale behaviour of materials (see \citep{vlassis2021sobolev, masi2021thermodynamics,masi2022multiscale,bonatti2022importance,as2023mechanics,cueto2023thermodynamics,vlassis2023geometric,rosenkranz2024viscoelasticty,holthusen2024theory,flaschel2025convex,qiu2025bridging,bleyer2025learning,fuhg2025review}, among others). Here, the second law of thermodynamics \citep{de2013non} plays a central role in identifying admissible evolution equations and has been enforced through both soft \citep{masi2021thermodynamics} and hard \citep{flaschel2025convex} constraints. In particular, data-driven methods based on (convex) dissipation (pseudo-)potentials provide a principled route to enforce non-negative dissipation by construction and have shown improved robustness and generalisation within their scope \citep{huang2022variational,as2023mechanics,rosenkranz2024viscoelasticty,holthusen2024theory,flaschel2025convex,holthusen2025generalized}. Closely related variational formulations have also been proposed to learn internal variables and coarse-grained dynamics from microscopic data with thermodynamic structure \citep{qiu2025bridging}. In parallel, structure-preserving (metriplectic) neural networks based on the GENERIC formalism \citep{ottinger2005beyond} have been proposed to enforce the first and second laws of thermodynamics via hard constraints within a broader non-equilibrium thermodynamic structure \citep{hernandez2021deep,cueto2023thermodynamics}.

Despite the aforementioned developments, data-driven approaches informed by physics often remain tied to specific constitutive classes or enforce only part of the relevant physical structure. In particular, formulations based on dissipation potentials provide a powerful route to thermodynamic consistency with hard constraints. However, they rely on the existence of a potential generating the underlying force--flux relation. This structural assumption may be restrictive for some admissible evolution laws \cite{van2008internal}, including non-associative and more general non-potential inelastic responses.

\begin{figure}[ht]
\centering
\includegraphics[width=0.85\textwidth]{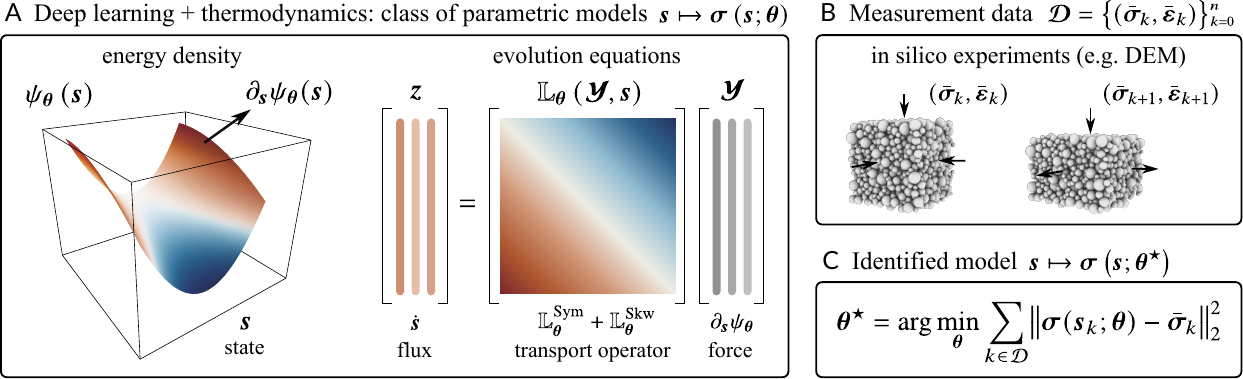}
\caption{Learning constitutive models from stress--strain data with hard-constrained non-equilibrium thermodynamics. (\textsf{A}) A thermodynamically admissible constitutive class $\bm{s}\mapsto \bm{\sigma}(\bm{s};\bm{\theta})$, parametrised by the state space $\bm{s}$, is built from two neural parametrisations: a free energy $\psi_{\bm{\theta}}(\bm{s})$, yielding thermodynamic forces $\partial_{\bm{s}}\psi_{\bm{\theta}}$ (first law), and a transport operator $\bm{\mathbb{L}}_{\bm{\theta}}(\bm{\mathcal{Y}}(\bm{s}),\bm{s})$ governing the evolution of $\bm{s}$ while enforcing non-negative dissipation (second law of thermodynamics). The networks are defined by a set of trainable parameters $\bm{\theta}$. The parametric model is trained against (\textsf{B}) discrete-time measurements $\smash{\mathcal{D}}=\bigl\{(\bar{\bm{\sigma}}_k,\bar{\bm{\varepsilon}}_k)\bigr\}$ sampled from in silico experiments as stress--strain pairs to identify (\textsf{C}) a ``best-fit'' parametrisation $\bm{\theta}^{\star}$ and the resulting constitutive model $\bm{s}\mapsto \bm{\sigma}(\bm{s}; \bm{\theta}^{\star})$.}
\label{fig:introduction}
\end{figure}

The aim of this work is to retain the key advantages of hard thermodynamic constraints while broadening the class of inelastic constitutive models that can be learned from labelled macroscopic stress--strain data via generalised transport equations (Figure~\ref{fig:introduction}). Building on thermodynamics-based artificial neural networks \citep{masi2021thermodynamics,masi2023evolution}, objective and convex energy-potential formulations (cf. \citep{klein2022polyconvex,linden2023neural}), and integral formulations of the evolution equations \citep{masi2024neural}, the proposed approach introduces a learning architecture that defines a class of admissible constitutive models by construction. The model is built on a structured, parametric state space $\bm{s}$ and a set of trainable neural parameters $\bm{\theta}$ through two coupled neural parametrisations: a free energy $\psi_{\bm{\theta}}(\bm{s})$ and a state- and force-dependent transport operator $\boldsymbol{\mathbb{L}}_{\bm{\theta}}(\bm{\mathcal{Y}},\bm{s})$ governing the evolution of $\bm{s}$, as depicted in Figure~\ref{fig:introduction}. The second law is enforced exactly through generalised transport equations \citep{gurtin1996generalized,van2008internal} rooted in non-equilibrium thermodynamics \citep{de2013non}. This formalism generalises Onsager--Casimir reciprocity relations \citep{onsagerI,onsagerII,casimir} and provides a systematic decomposition of reversible and irreversible contributions to the state evolution, without requiring an underlying dissipation-potential structure. The model parameters $\bm{\theta}^{\star}$ are identified by minimising the mismatch between predicted and measured stresses, yielding a data-driven constitutive closure that remains thermodynamically admissible at inference (Figure~\ref{fig:introduction}).

The performance and capabilities of the method are assessed on synthetic measurement datasets generated from analytical material models of increasing complexity, including elasto- and hypo-plasticity, rate effects, hardening, and non-associative flow rules. The identified models are shown to generalise to unobserved loading paths, and automatically identify interpretable internal variables, where needed, and consistent thermodynamic quantities (free energy and dissipation rate) even when these are not provided as measurements. Appendix~\ref{app:hard_soft_dissipation} demonstrates the benefits of the present framework against penalty-based (soft) formulations of the dissipation inequality.\\
The approach is then demonstrated on dry, cohesionless granular media using in silico experiments based on the discrete element method \citep{cundall1979discrete,kozicki2009yade}, with particular attention to aleatoric uncertainty arising from spatial heterogeneity and epistemic uncertainty due to limited observability of microstructural effects and states. Beyond identifying a reliable constitutive model over the observed loading protocols, the method extrapolates to unseen cyclic loading paths, including the emergence of hysteresis absent from the training data, while remaining microstructure-agnostic.\\

The paper is structured to first unfold the rationale behind the data-driven framework presented in Figure~\ref{fig:introduction}, and then to provide a rigorous assessment of its performance. Section~\ref{sec:theory} outlines the theoretical setting of non-equilibrium thermodynamics used throughout the work. Section~\ref{sec:methodology} presents the proposed data-driven framework, with particular emphasis on how hard constraints are enforced by construction in the neural parameterisations of the specific free energy and the transport operator, including objectivity, energy balance, and dissipation inequality. The resulting class of constitutive equations is then assembled and solved as an initial-value problem at the material-point level, and the training procedure is detailed. Section~\ref{sec:benchmarks} assesses accuracy, robustness, and generalisation on analytical material laws. Section~\ref{sec:discovery} applies the framework to granular media. Section~\ref{sec:conclusions} concludes with the main findings and perspectives.

\paragraph{Notation}
Scalar quantities are denoted by italic symbols $(\psi)$, first-order tensors by bold italic symbols $(\bm{v})$, second-order tensors by lowercase bold symbols $(\bm{\sigma})$, and higher-order tensors or generalised vectors by uppercase bold symbols $(\mathbf{L}, \bm{\mathcal{A}})$. Operator-valued quantities, such as the transport operator, are denoted by blackboard bold symbols $(\bm{\mathbb{L}})$.

Given arbitrary vectors ($\bm a,\bm b$) and second-order tensors ($\bm\sigma,\dot{\bm\varepsilon}$), the inner and outer products are denoted by $\smash{\bm a\cdot\bm b = a_i b_i}$, $\smash{\bm\sigma:\dot{\bm\varepsilon} = \sigma_{ij}\,\dot{\varepsilon}_{ij}}$, and $\smash{(\bm a \otimes \bm b)_{ij} = a_i b_j}$,  where summation over repeated indices is implied. The generalised inner product between two generalised vectors, $\bm{A}$ and $\bm{B}$, collecting quantities of possibly different tensorial order, is denoted by $\bm{A}\bm{\cdot}\bm{B}$ and is defined component-wise. For example, if $\bm{A}\equiv \left( \bm{a}\quad \bm{\sigma}\right)$ and $\bm{B}\equiv \left( \bm{b}\quad \dot{\bm{\varepsilon}}\right)$, then $\bm{A}\bm{\cdot} \bm{B} \equiv a_i b_i + \sigma_{ij} \dot{\varepsilon}_{ij}$. 

Spatial gradients of scalar and vector fields and the divergence of vector fields are defined by $\smash{(\nabla a)_i = \frac{\partial a}{\partial x_i}}$, $\smash{(\nabla \bm v)_{ij}= \frac{\partial v_i}{\partial x_j}}$, and $\smash{\operatorname{div}\bm{v} =  \frac{\partial v_i}{\partial x_i}}$. Superposed dots denote material time derivatives, $\mathbf{I}$ is the identity tensor, $\operatorname{tr}(\bm \sigma)=\sigma_{ii}$ denotes the trace, and the superscript $\top$ identifies the transpose operation.

\section{Non-equilibrium thermodynamics setting}
\label{sec:theory}
Consider a Cauchy continuum undergoing infinitesimal strains. Thermodynamic processes must comply with (\emph{i}) the mass balance, (\emph{ii}) the momentum balance, (\emph{iii}) the energy balance, (\emph{iv}) the entropy balance and second law of thermodynamics, and (\emph{v}) the principles of determinism, material frame-indifference (objectivity) and symmetry, among others. The pointwise balance of mass, linear momentum, and internal energy read 
\begin{equation}
\dot{\rho} +\rho \operatorname{div} \bm{v} = 0,
\end{equation}
\begin{equation}
    \rho \dot{\bm{v}} - \operatorname{div} \bm{\sigma} = \rho \bm{b},
\end{equation}
\begin{equation}
    \rho \dot{e} + \operatorname{div} \bm{q} = \bm{\sigma} \!:\! \nabla\bm{v}  + \rho r,
\end{equation}
where the latter two expressions are obtained using the mass balance. Here, $\rho$ is the mass density, $e$ is the specific internal energy (per unit mass), $\bm{v}$ is the velocity field, $\bm{\sigma}$ is the Cauchy stress, $\bm{b}$ is the body force density, $\bm{q}$ is the heat flux, and $r$ is the specific energy source. The balance of angular momentum implies the symmetry of the stress, i.e., $\bm{\sigma}=\bm{\sigma}^{\top}$.

The local entropy balance reads \citep{coleman1967thermodynamics}
\begin{equation}
\rho \dot{s} + \operatorname{div}\bm{k} = r_s + \gamma,
\end{equation}
where $s$ is the specific entropy, $T$ is the absolute temperature, $\bm{k}$ and $r_s$ are the entropy flux and supply, respectively, and $\gamma$ is the entropy production. The second law of thermodynamics requires $\gamma \geq 0$.
In the following, the canonical expressions for the entropy flux and supply, i.e., $\bm{k} \equiv T^{-1}\bm{q}$ and $r_s = \rho T^{-1} r$, are adopted. Note that the entropy flux could alternatively be regarded as a constitutive quantity itself to account for higher-grade materials (see \citep{maugin1990internal,ireman2004using,papenfuss2006thermodynamical,van2008internal}).

Combining the energy balance with the entropy inequality and introducing the specific Helmholtz free energy through the Legendre transformation $\psi = e-Ts$ yields the Clausius--Duhem inequality,
\begin{equation}
     T\gamma = \bm{\sigma}\!:\!\dot{\bm{\varepsilon}} + \rho \big[T\dot{s} - \dot{e}\big] - \frac{1}{T}\bm{q}\cdot \nabla T = \bm{\sigma}\!:\!\dot{\bm{\varepsilon}} - \rho \big [ \dot{\psi} + \dot{T}s\big ] -\frac{1}{T}\bm{q}\cdot\nabla T\geq 0,
\end{equation}
where $\dot{\bm{\varepsilon}}=\operatorname{Sym}(\nabla \bm{v})$ is the (infinitesimal) strain rate. In the following, the framework is restricted to isothermal processes ($\bm{q}=\boldsymbol{0}$, ${T} = \textrm{const}$) and neglects any covariant derivative (non-local features) to express the thermodynamic constraints in their simplest form. The Clausius--Duhem inequality reduces to 
\begin{equation}
    d \equiv T\gamma = \bm{\sigma}\!:\!\dot{\bm{\varepsilon}} - \rho \dot{\psi} \geq 0,
    \label{eq:clausius_}
\end{equation}
where $d$ is the internal dissipation rate.

\subsection{Constitutive restrictions and state laws}
Following \citet{coleman1967thermodynamics}, the state space of a material point $\mathbf{x}$ is described by a set of controllable state variables, such as the (elastic) strain and an additional set of internal (state) variables that encode the history-dependent state of the material and are non-controllable, i.e., their value cannot be prescribed through boundary conditions \citep{maugin1990internal}. Herein, the state of the material is assumed to be defined by  
\begin{equation}
\bm{s}=\begin{pmatrix} \rho & \bm{\varepsilon}^{{e}}& \bm{\alpha}_1 & \ldots & \bm{\alpha}_{n_{\bm{\alpha}}}\end{pmatrix},
\label{eq:state_space}
\end{equation}
where $\bm{\varepsilon}^{{e}}$ is the elastic strain and $\bm{\alpha}_i$, with $1\leq i \leq n_{\bm{\alpha}}$, are the $n_{\bm{\alpha}}$ internal variables. For compactness of notation, the set of internal variables is denoted by the generalised vector $\bm{\alpha} \equiv \begin{pmatrix}\bm{\alpha}_1 & \ldots & \bm{\alpha}_{n_{\bm{\alpha}}}\end{pmatrix}$.

Note that assuming $\bm{\varepsilon}^e$ to be a state variable does not imply that the underlying material behaviour is (elasto-)plastic (see Section~\ref{sec:benchmarks}) and it is a modelling choice that improves interpretability and often stabilises optimisation. An alternative is to parameterise the state in terms of the total strain $\bm{\varepsilon}$, which avoids an explicit decomposition of the strain rate into elastic and irreversible parts, but can make the state representation depend on the (arbitrarily chosen) reference configuration (see \citep{rubin1994plasticity}). Moreover, in experiments, strains are typically measured incrementally and reported relative to protocol-dependent reference states, thus using $\bm{\varepsilon}$ can introduce sensitivity to arbitrary offsets across tests. For completeness, Appendix~\ref{appendix:strain} presents the thermodynamic formulation obtained with a total-strain parametrisation. Note that the state space may also include higher-order components, e.g. $\nabla \rho, \nabla \bm{\varepsilon}^e, \nabla \bm{\alpha}$ \citep[cf.][]{papenfuss2006thermodynamical}. Such an extension is fully compatible with the present framework. However, these additional dependencies are omitted to keep the derivation of the transport relations hereafter as simple as possible.

By virtue of the classification of the state space~\eqref{eq:state_space}, the free energy is specified as
\begin{equation}
\psi \equiv \psi\left(\rho,\bm{\varepsilon}^{{e}},\bm{\alpha}\right).
\label{eq:energy}
\end{equation}

The thermodynamic forces conjugate to the state and internal variables are defined as
\begin{equation}
\mu \equiv \rho \frac{\partial \psi}{\partial \rho}\left(\rho,\bm{\varepsilon}^{{e}},\bm{\alpha}\right), \quad \bm{\sigma}^{{e}} \equiv \rho \frac{\partial \psi}{\partial \bm{\varepsilon}^{{e}}}\left(\rho,\bm{\varepsilon}^{{e}},\bm{\alpha}\right), \quad \bm{\mathcal{A}}^{{e}} \equiv \rho \frac{\partial \psi}{\partial \bm{\alpha}}\left(\rho,\bm{\varepsilon}^{{e}},\bm{\alpha}\right)
\label{eq:conj}
\end{equation}
with $\mu$ the chemical potential, $\bm{\sigma}^{{e}}$ the energetic stress, and $\bm{\mathcal{A}}^{{e}}$ the energetic forces associated with the internal variables. 
Assuming the free energy to be differentiable with respect to the state variables, its time derivative reads
\begin{equation}
\rho \dot{\psi} = \bm{\sigma}^{{e}} \!:\!\dot{\bm{\varepsilon}}^e + \mu \dot{\rho}+ \bm{\mathcal{A}}^{{e}}\bm{\cdot} \dot{\bm{\alpha}} = \bm{\sigma}^{{e}} \!:\!\big[\dot{\bm{\varepsilon}}-\dot{\bm{\varepsilon}}^{{p}} \big]-\rho{\mu}\operatorname{ tr}(\dot{\bm{\varepsilon}})+ \bm{\mathcal{A}}^{{e}}\bm{\cdot}\dot{\bm{\alpha}},
\label{eq:energy_rate}
\end{equation}
where the last expression is obtained by defining the plastic strain rate $\dot{\bm{\varepsilon}}^{{p}}\equiv \dot{\bm{\varepsilon}}-\dot{\bm{\varepsilon}}^{{e}}$ and using the mass balance, i.e., $\dot{\rho} = -\rho \operatorname{tr}(\dot{\bm{\varepsilon}})$.

\noindent Substitution of the time derivative~\eqref{eq:energy_rate} in the Clausius--Duhem inequality~\eqref{eq:clausius_} yields
\begin{equation}
    d = \bm{\sigma}^{{e}} \!:\!\dot{\bm{\varepsilon}}^{{p}}  + \dot{\bm{\varepsilon}}\!:\!\big[\bm{\sigma}-\bm{\sigma}^{{e}} +\rho{\mu}\mathbf{I} \big] -\bm{\mathcal{A}}^{{e}} \bm{\cdot} \dot{\bm{\alpha}} =  \bm{\sigma}^{{e}} \!:\!\dot{\bm{\varepsilon}}^{{p}} + \bm{\sigma}^{{d}}\!:\!\dot{\bm{\varepsilon}} +\bm{\mathcal{A}}^{{d}} \bm{\cdot} \dot{\bm{\alpha}} \geq 0,
    \label{eq:dissipation_full}
\end{equation}
where the dissipative stress $\bm{\sigma}^{{d}}$ and the dissipative driving forces $\bm{\mathcal{A}}^{{d}}$ are defined as
\begin{equation}
    \bm{\sigma}^{{d}} \equiv \bm{\sigma} -  \big[\bm{\sigma}^{{e}}\left(\rho,\bm{\varepsilon}^{{e}},\bm{\alpha}\right)-\rho{\mu}\left(\rho,\bm{\varepsilon}^{{e}},\bm{\alpha}\right)\mathbf{I}\big], \qquad \bm{\mathcal{A}}^{{d}} \equiv - \bm{\mathcal{A}}^{{e}}\left(\rho,\bm{\varepsilon}^{{e}},\bm{\alpha}\right).
    \label{eq:dissipative_forces}
\end{equation}
The dissipation inequality can thus be recast in the compact bilinear form
\begin{equation}
 d =  \bm{\mathcal{Y}} \bm{\cdot} \bm{z} \geq 0 ,\qquad  \bm{\mathcal{Y}} \equiv \begin{pmatrix}
    \bm{\sigma}^{{e}} &
    \bm{\mathcal{A}}^{{d}} &
    \bm{\sigma}^{{d}}
    \end{pmatrix}^{\top},\quad {\bm{z}} \equiv 
    \begin{pmatrix}
    \dot{\bm{\varepsilon}}^{{p}} &
    \dot{\bm{\alpha}}&
    \dot{\bm{\varepsilon}}
    \end{pmatrix}^{\top},
    \label{eq:dissipation}
\end{equation}
where $\bm{\mathcal{Y}}$ and $\bm{z}$ denote the generalised forces and generalised fluxes, respectively.

The next step is to specify an admissible force--flux relation $\bm{z}=\bm{g}(\bm{\mathcal{Y}},\bm{s})$, that is the plastic strain rate $\dot{\bm{\varepsilon}}^p$, the rate of the internal variables $\dot{\bm{\alpha}}$, and the total strain rate $\dot{\bm{\varepsilon}}$ (or the dissipative stress $\bm{\sigma}^d$), such that $\bm{\mathcal{Y}}\bm{\cdot} \bm{g}(\bm{\mathcal{Y}},\bm{s}) \geq 0$, for all admissible $(\bm{\mathcal{Y}},\bm{s})$ according to the state laws \eqref{eq:conj} and Eq.~\eqref{eq:dissipative_forces}, see also \citep{van2008internal,einav2018hydrodynamic}.

\subsection{Transport equations}
\label{subsec:transport}
A broad class of solutions of the dissipation inequality~\eqref{eq:dissipation} is provided by the generalised transport equations \citep{gurtin1996generalized,van2008internal}
\begin{equation}
\bm{z} = \bm{\mathbb{L}} \big(\bm{\mathcal{Y}}(\bm{s}),\bm{s}\big) \bm{\mathcal{Y}},
    \quad \text{i.e.} \quad 
    \begin{pmatrix}
    \dot{\bm{\varepsilon}}^{{p}}\\
    \dot{\bm{\alpha}}\\
    \dot{\bm{\varepsilon}}
    \end{pmatrix}=
    \begin{pmatrix}
    \mathbf{L}^{{11}} & \mathbf{L}^{{12}} & \mathbf{L}^{{13}} \\
    \mathbf{L}^{{21}} & \mathbf{L}^{{22}} & \mathbf{L}^{{23}}\\
    \mathbf{L}^{{31}} & \mathbf{L}^{{32}} & \mathbf{L}^{{33}}
    \end{pmatrix}
    \begin{pmatrix}
    \bm{\sigma}^{{e}} \\
    \bm{\mathcal{A}}^{{d}}\\
    \bm{\sigma}^{{d}}
    \end{pmatrix}
    \label{eq:L_first}
\end{equation}
where the blocks $\mathbf{L}^{ij}$ of the transport operator $\bm{\mathbb{L}}$ are in general nonlinear operators mapping the $j$th force component to the $i$th flux component, and may depend on the state variables and the generalised forces, $\mathbf{L}^{ij}=\mathbf{L}^{ij}\big(\bm{\mathcal{Y}}(\bm{s}),\bm{s}\big)$, along with additional variables, when required.

By decomposing the transport operator into symmetric and skew-symmetric parts, $\operatorname{Sym}(\bm{\mathbb{L}})$ and $\operatorname{Skw}(\bm{\mathbb{L}})$, defined with respect to the generalised inner product $\bm{\cdot}$, the evolution equations read
\begin{equation}
\begin{split}
    \bm{z} &= \operatorname{Sym}(\bm{\mathbb{L}})\bm{\mathcal{Y}}
           + \operatorname{Skw}(\bm{\mathbb{L}})\bm{\mathcal{Y}} 
           \\ 
    \begin{pmatrix}
    \dot{\bm{\varepsilon}}^{{p}}\\
    \dot{\bm{\alpha}}\\
    \dot{\bm{\varepsilon}}
    \end{pmatrix}& =\underbrace{\frac{1}{2}
    \begin{pmatrix}
    2\mathbf{L}^{{11}} & \mathbf{L}^{{12}}+\mathbf{L}^{{21}} & \mathbf{L}^{{13}}+\mathbf{L}^{{31}} \\
    \mathbf{L}^{{21}}+\mathbf{L}^{{12}} & 2\mathbf{L}^{{22}} & \mathbf{L}^{{23}}+\mathbf{L}^{{32}}\\
    \mathbf{L}^{{31}}+\mathbf{L}^{{13}}& \mathbf{L}^{{32}}+\mathbf{L}^{{23}} & 2\mathbf{L}^{{33}}
    \end{pmatrix}}_{\displaystyle \operatorname{Sym}(\bm{\mathbb{{L}}})}
    \begin{pmatrix}
    \bm{\sigma}^{{e}} \\
    \bm{\mathcal{A}}^{{d}}\\
    \bm{\sigma}^{{d}}
    \end{pmatrix} +\underbrace{\frac{1}{2}
    \begin{pmatrix}
    \boldsymbol{0}& \mathbf{L}^{{12}}-\mathbf{L}^{{21}} & \mathbf{L}^{{13}}-\mathbf{L}^{{31}} \\
    \mathbf{L}^{{21}}-\mathbf{L}^{{12}} & \boldsymbol{0} & \mathbf{L}^{{23}}-\mathbf{L}^{{32}}\\
    \mathbf{L}^{{31}}-\mathbf{L}^{{13}}& \mathbf{L}^{{32}}-\mathbf{L}^{{23}} & \boldsymbol{0}
    \end{pmatrix}}_{\displaystyle \operatorname{Skw}(\bm{\mathbb{{L}}})}
    \begin{pmatrix}
    \bm{\sigma}^{{e}} \\
    \bm{\mathcal{A}}^{{d}}\\
    \bm{\sigma}^{{d}}
    \end{pmatrix}.
    \end{split}
\end{equation}
The dissipation inequality thus becomes
\begin{equation}
    d = \bm{\mathcal{Y}}\bm{\cdot}\bm{z}
      = \bm{\mathcal{Y}}\bm{\cdot}\big(\operatorname{Sym}(\bm{\mathbb{L}})\bm{\mathcal{Y}}\big)
      + \bm{\mathcal{Y}}\bm{\cdot}\big(\operatorname{Skw}(\bm{\mathbb{L}})\bm{\mathcal{Y}}\big) = \bm{\mathcal{Y}}\bm{\cdot}\big(\operatorname{Sym}(\bm{\mathbb{L}})\bm{\mathcal{Y}}\big),
\end{equation}
because skew-symmetry implies $\bm{\mathcal{Y}}\bm{\cdot}\big(\operatorname{Skw}(\bm{\mathbb{L}})\bm{\mathcal{Y}}\big)=0$ for all $\bm{\mathcal{Y}}$. This yields a decomposition of the evolution equations, driven by the thermodynamic forces derived from $\psi$, into dissipative and non-dissipative contributions generated respectively by its symmetric and skew-symmetric parts. The latter contribution is often referred to as gyroscopic \citep{van2008internal} and accounts for possible reversible couplings in the material state evolution.

Finally, a sufficient condition for the dissipation inequality $d \ge 0$ to hold for all admissible generalised forces and the corresponding fluxes is that the symmetric part of the transport operator is positive semidefinite, i.e.,
\begin{equation}
    \operatorname{Sym}(\bm{\mathbb{L}}) \succeq 0 \quad \text{for all admissible }(\bm{\mathcal Y},\bm s).
\end{equation}
This condition ensures thermodynamic admissibility of the force--flux relation by constraining the sign of the dissipation rate. Note, however, that it does not prescribe the rate character of the material response. Rate independence is an additional constitutive property, requiring invariance under monotone reparametrisations of time \citep{mielke2005evolution}. In the present transport setting, it can be imposed, when required, by building evolution laws that are positively homogeneous of degree one with respect to the imposed rates, for instance through a dependence of the transport operator on loading direction but not on loading-rate magnitude. 

It is also worth noticing that the present framework is intrinsically linked with the extension of non-equilibrium thermodynamics with dual internal variables, proposed by \citet{van2008internal}. Although non-local terms (e.g. $\nabla\bm\alpha$) are omitted here, the set $(\bm{\alpha}_1,\ldots,\bm{\alpha}_{n_{\bm{\alpha}}})$ is not fixed a priori and may include dual internal variables. The dual-variables approach unifies the thermodynamic role of (dissipative) internal variables with internal degrees of freedom that possess their own balance equations, enabling the inclusion of micro-inertia effects \citep{berezovski2016microinertia} and the tracking of internal dynamics \citep{van2008internal,berezovski2018internal}, without the more stringent requirements assumed in the GENERIC framework \citep{{ottinger2005beyond}}, e.g. the degeneracy conditions.

\subsubsection{Dissipation potentials and the standard materials theory}
\label{subsubsec:potential}
An alternative path to identify thermodynamically admissible fluxes consists of resorting to dissipation potentials \citep{ziegler1958attempt,moreau1970lois,Moreau2011} and the theory of Generalised Standard Materials (GSM \citep{halphen1975materiaux,mcbride2018dissipation}). By postulating the existence of a thermodynamic potential $\phi\left(\bm{\mathcal{Y}}\right)$, convex and lower semicontinuous, such that the fluxes $\bm{z}$ associated with the forces $\bm{\mathcal{Y}}$ are the gradients of $\phi$, i.e., $\bm{z}\equiv \partial \phi/ \partial \bm{\mathcal{Y}}$, with the requirement that $ \partial \phi \left(\boldsymbol{0}\right) / \partial \bm{\mathcal{Y}}=\boldsymbol{0}$, the fulfilment of the dissipation inequality is ensured by construction, i.e., $d=\bm{\mathcal{Y}}\bm{\cdot} \partial \phi/ \partial \bm{\mathcal{Y}}\geq 0$. The same formalism has been extended to constitutive classes not covered by classical GSM (e.g. non-associative plasticity) by introducing implicit representations of (bi-)potentials, cf.\ the implicit standard materials framework \citep{de1991new,deSaxce2002}.

Despite the many successful applications to inelastic and complex material responses \citep{hyperplasticity,mcbride2018dissipation,huang2022variational,as2023mechanics,rosenkranz2024viscoelasticty,holthusen2024theory,flaschel2025convex,holthusen2025generalized}, including data-driven approaches, dissipation-potential approaches rely on the a priori existence of such a potential and therefore impose stronger structural requirements than those adopted here. In particular, under sufficient smoothness of the force--flux map $\bm{\mathcal{Y}}\mapsto \bm{z}$, the existence of a dissipation potential requires the Jacobian to be symmetric, i.e., $\partial \bm{z}/\partial \bm{\mathcal{Y}}=\left(\partial \bm{z}/\partial \bm{\mathcal{Y}}\right)^{\top}$. For the generalised transport equations~\eqref{eq:L_first}, this Jacobian contains not only the transport operator $\bm{\mathbb L}(\bm{\mathcal Y},\bm s)$, but also additional terms arising from the possible dependence of $\bm{\mathbb L}$ on the generalised forces. Therefore, even when $\bm{\mathbb L}=\bm{\mathbb L}^{\top}\succeq0$ pointwise, the Jacobian $\partial\bm z/\partial\bm{\mathcal Y}$ need not be symmetric if the transport operator is force-dependent. Thus, symmetry and positive semidefiniteness of $\bm{\mathbb L}$ do not imply the existence of a scalar dissipation potential (for more details, see \citep{van2008internal}). At the same time, generalised transport equations parameterise vector- or tensor-valued operators, whereas dissipation-potential formulations generate the evolution law from scalar potentials and therefore involve fewer unknown constitutive maps.

\subsubsection{Onsager--Casimir reciprocity relations} Under the assumption of time-reversal symmetry of the evolution equations, the generalised transport coefficients in Eq.~\eqref{eq:L_first} must satisfy the Onsager--Casimir reciprocity relations, i.e., $\mathbf{L}^{21}=\mathbf{L}^{12}$, $\mathbf{L}^{31}=-\mathbf{L}^{13}$, $\mathbf{L}^{32}=-\mathbf{L}^{23}$,
where the sign depends on the parity of the associated force--flux pairs under time reversal \citep{onsagerI,onsagerII,casimir,de2013non}. These relations were derived in statistical physics from the principle of microscopic reversibility and have since been successfully adopted in the constitutive modelling of inelastic materials (among others, see \citep{jiang2009granular,jiang2015applying,einav2018hydrodynamic,einav2023hydrodynamics,wiebicke2024simple,masi2025hydrodynamics}). However, when the state space includes arbitrary internal variables whose microscopic counterparts and time-reversal parities are unknown, imposing reciprocity may be too restrictive and not always appropriate. This is particularly true when there is no a priori phenomenological model establishing the physical nature of each state variable. On that basis, the present framework does not enforce the Onsager--Casimir reciprocity relations, by construction. Nevertheless, whenever the internal variables are measurable, their parities are established, and can be linked to microscopic phenomena, reciprocity constraints can be added without otherwise modifying the approach.\\

\noindent For simplicity, the dissipative stress is neglected henceforth, as it is commonly assumed in solid mechanics. The transport equations therefore reduce to
\begin{equation}
    \bm{z}= \bm{\mathbb{L}} \big(\bm{\mathcal{Y}}(\bm{s}),\bm{s}\big) \bm{\mathcal{Y}},
    \quad  \bm{\mathcal{Y}} \equiv \begin{pmatrix}
    \bm{\sigma}^{{e}} &
    \bm{\mathcal{A}}^{{d}}
    \end{pmatrix}^{\top},\quad {\bm{z}} \equiv 
    \begin{pmatrix}
    \dot{\bm{\varepsilon}}^{{p}} &
    \dot{\bm{\alpha}}
    \end{pmatrix}^{\top},
    \qquad \text{i.e.} \quad 
    \begin{pmatrix}
    \dot{\bm{\varepsilon}}^{{p}}\\
    \dot{\bm{\alpha}}    
    \end{pmatrix}=
    \begin{pmatrix}
    \mathbf{L}^{{11}} & \mathbf{L}^{{12}} \\
    \mathbf{L}^{{21}} & \mathbf{L}^{{22}} \\
    \end{pmatrix}
    \begin{pmatrix}
    \bm{\sigma}^{{e}} \\
    \bm{\mathcal{A}}^{{d}}
    \end{pmatrix}.
    \label{eq:transport_eqs_final}
\end{equation}

\section{Thermodynamics-constrained constitutive learning}
\label{sec:methodology}
Building on the above non-equilibrium thermodynamic setting, the goal is to identify the specific free energy $\psi$ and the transport operator $\bm{\mathbb L}$ so that the resulting constitutive equations fulfil the thermodynamic principles by construction. To this end, two general neural-network ansatz functions are parametrised with respect to the set of neural network parameters $\bm{\theta}$ for the two unknown operators, namely
\begin{equation}
    \psi_{\bm{\theta}} \left(\bm{s} \right)\approx \psi \left(\bm{s} \right), \qquad \bm{\mathbb{L}}_{\bm{\theta}}\left( \bm{\mathcal{Y}}(\bm{s}), \bm{s}\right) \approx \bm{\mathbb{L}}\left( \bm{\mathcal{Y}}(\bm{s}), \bm{s}\right),
    \label{eq:ansatz}
\end{equation} 
where $\bm s$ denotes the state space and $\bm{\mathcal Y}$ the generalised forces.

\subsection{Hard constraints}
\label{subsec:hard_constraints}
As detailed in Section~\ref{sec:theory}, the ansatz functions in \eqref{eq:ansatz} are required to satisfy: (\emph{i}) the state laws \eqref{eq:dissipative_forces}; (\emph{ii}) the positive semidefiniteness of the symmetric part of the transport operator; and (\emph{iii}) the frame-indifference principle. The following postulates are additionally considered: (\emph{iv}) convexity constraints on the free energy and (\emph{v}) vanishing energy and conjugate forces at an equilibrium stress-free configuration. Item~(\emph{iv}) promotes volumetric and elastic stability of the energy response with respect to the selected invariant representation, as detailed in Section~\ref{subsubsec:energy_net}. Note however that this condition does not preclude loss of stability (e.g. softening) associated with the evolution equations and internal-variable dynamics. Item (\emph{v}) ensures the existence of a reference configuration $\bm{s}_{\mathrm{ref}}$ -- undeformed ($\bm{\varepsilon}^e_{\mathrm{ref}} =\boldsymbol{0}$), at finite density $\rho_{\mathrm{ref}}>0$ and internal state $\bm{\alpha}_{\mathrm{ref}}=\boldsymbol{0}$ -- which is stress-free and has zero stored energy. Depending on the application, convexity with respect to internal variables $\bm\alpha$ may also be considered (e.g. \citep{flaschel2025convex}). In the following, individual internal variables $\bm\alpha_i$, $1\leq i \leq n_{\bm \alpha}$, are assumed scalar.

Frame indifference is enforced by formulating the constitutive response in objective coordinates. The framework assumes an isotropic response and expresses the relevant tensorial quantities through scalar invariants. Let
\begin{equation}
    \bm s_I = \begin{pmatrix}
        \rho & \bm\varepsilon_I^e & \bm\alpha 
    \end{pmatrix}^{\top}, \quad
    \bm{\mathcal{Y}}_I = \begin{pmatrix}
        \bm\sigma_I^e & \bm{\mathcal{A}}^d 
    \end{pmatrix}^{\top},  \quad
    \bm{z}_I = \begin{pmatrix}
        \dot{\bm{\varepsilon}}^p_I & \dot{\bm{\alpha}} 
    \end{pmatrix}^{\top}
\end{equation}
denote, respectively, the reduced state-space vector, generalised-force vector, and flux vector. Here, $\bm\varepsilon_I^e$ and $\dot{\bm\varepsilon}_I^p$ collect objective elastic-strain and plastic strain-rate coordinates, while $\bm\sigma_I^e$ collects the corresponding energetic-stress coordinates. The neural network ansatz functions in Eq.~\eqref{eq:ansatz} are therefore constructed and evaluated in this objective representation, i.e.,
\begin{equation}
    \hat{\psi}_{\bm\theta}(\bm s_I) \equiv \psi_{\bm\theta}(\bm s), \qquad \hat{\;\bm{\mathbb L}}_{\bm\theta}(\bm{\mathcal Y}_I,\bm s_I) \equiv {\bm{\mathbb L}}_{\bm\theta}(\bm{\mathcal Y},\bm s),
    \label{eq:invariant_maps}
\end{equation}
where hats denote the invariant-coordinate representation of the corresponding constitutive maps.

In the present setting, the stress and strain invariant vectors are chosen as
$\smash{\bm{\sigma}_I = \left(p \quad q \right)^{\top}}$ and $\smash{\bm{\varepsilon}_I=\left( \varepsilon_v \quad \varepsilon_s \right)^{\top}}$, with $p =  -1/3\operatorname{tr}(\bm{\sigma})$ and $\varepsilon_v = -\operatorname{tr}(\bm{\varepsilon})$, considering positive compressive stress and strain, and with $\smash{q = \sqrt{3/2\,\bm{\sigma}'\!:\!\bm{\sigma}'}}$ and $\smash{\varepsilon_s = \sqrt{2/3\,\bm{\varepsilon}'\!:\!\bm{\varepsilon}'}}$. Here, $\smash{\bm{\chi}'=\bm{\chi}-1/3\operatorname{tr}(\bm{\chi})\mathbf{I}}$ denotes the deviatoric part of an arbitrary second-order tensor $\bm{\chi}$.
 Note that additional invariant coordinates (e.g. third deviatoric invariants) may be appropriately included when required. Extensions to anisotropy and to non-coaxial stress--strain relations can be obtained by augmenting the state with structural tensors and by adopting equivariant architectures (cf. \citep{Heider2020SO3}). 

\subsubsection{Free-energy parametrisation}
\label{subsubsec:energy_net}

Convexity of the free energy with respect to $\rho$ and objective elastic-strain coordinates $\bm\varepsilon^{e}_I$ is enforced using a partially input-convex neural network \citep{amos2017input}, which guarantees convexity in the selected inputs through non-negative weights and convex, non-decreasing activations along the convex branch. The resulting energy is convex in the constrained inputs, whereas macroscopic nonlinearities and history effects remain governed by internal-variable dynamics and the transport equations. Vanishing energy and conjugate forces at the reference state $\bm{s}_{I,\mathrm{ref}}$ are enforced by affine correction terms, namely
\begin{equation}
    \hat{\psi}_{\bm{\theta}}(\bm{s}_{\!I})
    \equiv
    \hat{\psi}^{\mathrm{C}}_{\bm{\theta}}(\bm{s}_{\!I})
    - \hat{\psi}_{\bm{\theta}}^{\mathrm{C}}(\bm{s}_{\!I,\mathrm{ref}})
    - \frac{\partial \hat{\psi}_{\bm{\theta}}^{\mathrm{C}}}{\partial \bm{s}_{\!I}}(\bm{s}_{\!I,\mathrm{ref}})
    \bm{\cdot} \left(\bm{s}_{\!I}-\bm{s}_{\!I,\mathrm{ref}}\right),
    \label{eq:energy_correction}
\end{equation}
where $\hat{\psi}_{\bm{\theta}}^{\mathrm C}(\bm{s}_I)$ is a partially input-convex neural network fed with the invariant representations of the elastic strain, the density, and the internal variables. The reference density $\rho_{\mathrm{ref}}$ is treated as trainable network parameter (constrained to be positive), so that the network can identify the stress-free state directly from data in the selected invariant representation. This definition gives, by construction,
\begin{equation}
\hat{\psi}_{\bm\theta}(\bm{s}_{I,\mathrm{ref}})=0,\qquad
\frac{\partial \hat{\psi}_{\bm\theta}}{\partial \rho}(\bm{s}_{I,\mathrm{ref}})=0,\qquad
\frac{\partial \hat{\psi}_{\bm\theta}}{\partial \bm\varepsilon^{e}_I}(\bm{s}_{I,\mathrm{ref}})=\boldsymbol 0,\qquad
\frac{\partial \hat{\psi}_{\bm\theta}}{\partial \bm\alpha}(\bm{s}_{I,\mathrm{ref}})=\boldsymbol 0.
\end{equation}

The thermodynamic forces $\mu$ and $\bm{\mathcal{A}}^e$ and energetic stress invariants $\bm{\sigma}^e_{\!I}$ are computed by automatic differentiation of the free energy with respect to the reduced state-space vector following Eq.~\eqref{eq:conj}, namely
\begin{equation}
\mu = \rho \frac{\partial \hat{\psi}_{\bm\theta}}{\partial \rho}(\bm s_I),
\qquad
\bm\sigma_I^e = \rho\frac{\partial \hat{\psi}_{\bm\theta}}{\partial \bm\varepsilon_I^e}(\bm s_I), \qquad
\bm{\mathcal A}^e = \rho \frac{\partial \hat{\psi}_{\bm\theta}}
{\partial \bm\alpha}(\bm s_I),
\end{equation}
resulting in the fulfilment of the energy balance, see \cite{masi2021thermodynamics}. Finally, the stress invariants $\bm{\sigma}_{\!I}$ follow from the invariant representation of Eq.~\eqref{eq:dissipative_forces}, namely $\bm\sigma_I = \bm\sigma_I^e + \rho  \mu \mathbf e_v$, where $\smash{\mathbf e_v=(1 \quad 0)^{\top}}$ denotes the volumetric direction.

\subsubsection{Transport-operator parametrisation}
\label{subsubsec:evolution_net}
To enforce the dissipation inequality, the transport operator ansatz is split into symmetric and skew-symmetric parts and parametrised as
\begin{equation}
    \hat{\,\bm{\mathbb{L}}}_{\bm{\theta}} = \hat{\,\bm{\mathbb{L}}}_{\bm{\theta}}^{\textrm{Sym}}+\hat{\,\bm{\mathbb{L}}}_{\bm{\theta}}^{\textrm{Skw}},\qquad  
    \hat{\,\bm{\mathbb{L}}}_{\bm{\theta}}^{\textrm{Sym}} = \hat{\bm{\mathbb{T}}}_{\bm{\theta}}\hat{\bm{\mathbb{T}}}_{\bm{\theta}}^{\top}\succeq 0 , \qquad \hat{\,\bm{\mathbb{L}}}_{\bm{\theta}}^{\textrm{Skw}} = \begin{pmatrix} 
    \boldsymbol{0} & \bm{\mathscr{k}}_{\bm{\theta}}\vspace{2pt}\\
    -\bm{\mathscr{k}}^\top_{\bm{\theta}} & \boldsymbol{0}
    \end{pmatrix}, \qquad \hat{\bm{\mathbb{T}}}_{\bm{\theta}} = \begin{pmatrix} 
    \bm{\mathscr{l}}^{11}_{\bm{\theta}} & \boldsymbol{0}\vspace{2pt}\\
    \bm{\mathscr{l}}^{12}_{\bm{\theta}} & \bm{\mathscr{l}}^{22}_{\bm{\theta}}\end{pmatrix},
     \label{eq:L_param}
\end{equation}
where all sub-blocks $\smash{\bm{\mathscr{l}}^{11}_{\bm{\theta}},\bm{\mathscr{l}}^{12}_{\bm{\theta}},\bm{\mathscr{l}}^{22}_{\bm{\theta}},\bm{\mathscr{k}}_{\bm{\theta}}}$ are the outputs of a feed-forward neural network $\bm{f}_{\bm{\theta}}(\bm{\mathcal{Y}}_I,\bm{s}_I)$. Since $\smash{\hat{\;\bm{\mathbb{L}}}_{\bm{\theta}}^{\textrm{Sym}}}$ is expressed in terms of its Cholesky factorisation from the lower triangular matrix $\smash{\hat{\bm{\mathbb{T}}}_{\bm{\theta}}}$, it is positive semidefinite by construction and Eq.~\eqref{eq:L_param} can approximate transport operators while enforcing non-negative dissipation by construction.

The final form of the transport equations reads
\begin{equation}
\begin{pmatrix}
    \dot{\bm{\varepsilon}}^{{p}}_{\!I}\vspace{3pt}\\
    \dot{\bm{\alpha}}    
    \end{pmatrix}=
    \begin{pmatrix}
    {\mathbf{L}}_{\bm{\theta}}^{11}&{\mathbf{L}}_{\bm{\theta}}^{12}\vspace{3pt}\\
    {\mathbf{L}}_{\bm{\theta}}^{21} & {\mathbf{L}}_{\bm{\theta}}^{22}
    \end{pmatrix}
    \begin{pmatrix}
    \bm{\sigma}^{{e}}_{\!I} \vspace{3pt}\\
    \bm{\mathcal{A}}^{{d}}
    \end{pmatrix}
    \label{eq:transport_eqs_net}
\end{equation}
with
\begin{align}
    {\mathbf{L}}_{\bm{\theta}}^{11}&=\bm{\mathscr{l}}^{11}_{\bm{\theta}}
\left(\bm{\mathscr{l}}^{11}_{\bm{\theta}}\right)^\top, \qquad\qquad\,\,
\mathbf{L}^{12}_{\bm{\theta}}
= \bm{\mathscr{l}}^{11}_{\bm{\theta}}
\left(\bm{\mathscr{l}}^{12}_{\bm{\theta}}\right)^\top
+ \bm{\mathscr{k}}_{\bm{\theta}},\\
{\mathbf{L}}_{\bm{\theta}}^{21}&=\bm{\mathscr{l}}^{12}_{\bm{\theta}}
\left(\bm{\mathscr{l}}^{11}_{\bm{\theta}}\right)^\top
- \bm{\mathscr{k}}_{\bm{\theta}}^\top,\qquad
    {\mathbf{L}}_{\bm{\theta}}^{22}=\bm{\mathscr{l}}^{12}_{\bm{\theta}}
\left(\bm{\mathscr{l}}^{12}_{\bm{\theta}}\right)^\top
+
\bm{\mathscr{l}}^{22}_{\bm{\theta}}
\left(\bm{\mathscr{l}}^{22}_{\bm{\theta}}\right)^\top,
\end{align}
and $\bm{\mathcal{A}}^d = -\bm{\mathcal{A}}^e$. 
This structure enables dissipative and non-dissipative force--flux couplings, including non-associative evolution laws beyond generalised standard materials, cf.\ Section~\ref{subsubsec:potential}.

\subsection{Constitutive update and time integration}
The constitutive response under a given loading protocol is obtained by solving the following differential-algebraic system of equations, which couples the evolution equations with the state laws for the stress coordinates $\bm{\sigma}_I$ and the generalised-force coordinates $\bm{\mathcal{Y}_I}$,

\begin{equation}
\begin{cases}
    \;\dot{\rho} = \rho \dot{\varepsilon}_v\\
    \;\dot{\bm{\varepsilon}}^{ e}_{\!I} = \dot{\bm{\varepsilon}}_{\!I}-\rho\left[{\mathbf{L}}_{\bm{\theta}}^{11}\dfrac{\partial \hat{\psi}_{\bm{\theta}}}{\partial \bm{\varepsilon}^e_{\!I}}(\bm{s}_{\!I})-{\mathbf{L}}_{\bm{\theta}}^{12}\dfrac{\partial \hat{\psi}_{\bm{\theta}}}{\partial \bm{\alpha}}(\bm{s}_{\!I})\right]\vspace{1pt}\\
    \;\dot{\bm{\alpha}} = \rho\left[{\mathbf{L}}_{\bm{\theta}}^{21}\dfrac{\partial \hat{\psi}_{\bm{\theta}}}{\partial \bm{\varepsilon}^e_{\!I}}(\bm{s}_{\!I})- {\mathbf{L}}_{\bm{\theta}}^{22}\dfrac{\partial \hat{\psi}_{\bm{\theta}}}{\partial \bm{\alpha}}(\bm{s}_{\!I})\right]\vspace{3pt}\\
    \;\bm{\sigma}_{\!I} = \rho\left[ \dfrac{\partial \hat{\psi}_{\bm{\theta}}}{\partial \bm{\varepsilon}^e_{\!I}}(\bm{s}_{\!I}) +\rho\mathbf{e}_v\dfrac{\partial \hat{\psi}_{\bm{\theta}}}{\partial \rho}(\bm{s}_{\!I}) \right]\vspace{4pt}\\
    \;\bm{\mathcal{Y}}_{\!I} =  \begin{pmatrix}
    \rho\dfrac{\partial \hat{\psi}_{\bm{\theta}}}{\partial \bm{\varepsilon}^{e}_{\!I}}(\bm{s}_{\!I}) &
    -\rho\dfrac{\partial \hat{\psi}_{\bm{\theta}}}{\partial \bm{\alpha}}(\bm{s}_{\!I})
    \end{pmatrix}^{\top}.
    \label{eq:DAE}
\end{cases}
\end{equation}
In the remainder of this section, a displacement-driven setting with prescribed $\bm{\varepsilon}(t)$, or equivalently $\bm{\varepsilon}_I(t)$, is considered for simplicity. When a protocol requires (full or mixed) force control, the unknown strain-rate coordinates are recovered by augmenting Eq.~\eqref{eq:DAE} with the corresponding algebraic stress constraints (e.g. $\bm{\sigma}_{I}=\text{const}$) and solving the resulting nonlinear system by means of (quasi-)Newton iterations.

\subsubsection{Time integration and solution}
\label{subsubsec:solveivp}
The initial-value problem for the state and internal variables reads
\begin{equation}
    \bm{s}_I(t) = \bm{s}_{I,0}+ \int_{t_0}^t \dfrac{d \bm{s}_I}{d t}\big( \bm{s}_I(t), \dot{\bm{\varepsilon}}_I(t); \bm{\theta}\big)\, \text{d}t,
    \qquad \bm{s}_{I,0} = 
        \begin{pmatrix}
        \rho_0 & \bm{\varepsilon}^{e}_{I,0} & \bm{\alpha}_0 \end{pmatrix}^{\top},
\label{eq:ivp}
\end{equation}
where $\bm{s}_{I,0}$ denotes the initial state-space vector and $\frac{d \bm{s}_I}{d t}$ collects the rates of the state and the internal variables given by the right-hand sides of Eq.~\eqref{eq:DAE}, parametrised with respect to the network parameters $\bm{\theta}$.

The solution of Eq.~\eqref{eq:ivp} can be obtained using standard numerical integrators for ordinary differential equations, including explicit and implicit schemes (fixed-step or adaptive), and since the right-hand side is parametrised by neural networks, the initial-value problem can be interpreted as a neural differential equation \citep{chen2018neural,masi2024neural}.

Once the time-history of the material state $\bm{s}_{I,n}$ is computed, with $\bm s_{I,n}=\bm s_I(t_n)$ the state at time $t_n$, the evolution of the stress coordinates is directly obtained from the state laws, i.e.,
\begin{equation}
    {\bm{\sigma}}_{I,n}
    \equiv {\bm{\sigma}}_I\!\left(\bm s_{I,n};\bm\theta\right)
    =
    \rho_{n}\left[\frac{\partial \hat{\psi}_{\bm\theta}}{\partial \bm{\varepsilon}^{e}_I}\!\left(\bm s_{I,n}\right)
    +
    \rho_{n}\mathbf{e}_v\frac{\partial \hat{\psi}_{\bm\theta}}{\partial \rho}\!\left(\bm s_{I,n}\right)\right].
    \label{eq:stress_discrete}
\end{equation}

\subsection{Learning constitutive equations from data}
\label{subsec:training}
Building on the solution of the constitutive equations parameterised by $\hat{\psi}_{\bm{\theta}}$ and $\hat{\;\bm{\mathbb{L}}}_{\bm{\theta}}$ that are hard-constrained to satisfy the principles of non-equilibrium thermodynamics, the aim is to identify the network parameters $\bm{\theta}^{\star}$ that best fit measured responses under prescribed loading protocols. Measurements may originate from high-fidelity numerical simulations or laboratory experiments. In general, the specific free energy ($\psi$), the internal variables ($\bm{\alpha}$), the elastic strain ($\bm{\varepsilon}^e$) and their rates are not directly accessible, nor can they be inferred a priori without positing a specific constitutive model. The same holds for the thermodynamic conjugate variables ($\bm{\sigma}^e$, $\mu$, and $\bm{\mathcal{A}}^{e}$). Additionally, in practice, laboratory measurements are often recorded at sparse and irregular time intervals, which make the use of incremental formulations non-trivial (for more details, see \citep{masi2024neural}).

The appeal of the proposed approach is that, despite the above challenges, it can identify physically consistent internal variables and learn thermodynamically consistent constitutive equations (state laws and evolution equations) in the presence of sparse and incomplete measurements. In this work, attention is restricted to the common setting in which only labelled stress--strain pairs and the material density are available, see Figure \ref{fig:introduction}.

\subsubsection{Initial conditions}
\label{subsubsec:init_cond}
To predict the time evolution of the state variables, initial conditions for the material state are required. While the initial density is readily known or measured, the elastic strain and the internal variables are not directly accessible unless restrictive assumptions are made (e.g. a prescribed energy form).

Following \citep{masi2024neural} and with the aim of developing a scalable approach capable of handling the broadest range of initial conditions, the initial elastic-strain coordinates are obtained from the measured initial stress $\bar{\bm{\sigma}}_{I,0}$ and the measured initial density $\bar{\rho}_0$ by solving the nonlinear algebraic equation
\begin{equation}
    \bm{\mathcal{R}}\left(\bm{s}_{I,0}, \bar{\bm{\sigma}}_{I,0}; \bm{\theta}\right)
    = {\bm{\sigma}}_I\left( \bm{s}_{I,0}; \bm{\theta}\right) - \bar{\bm{\sigma}}_{I,0}
    = \bar{\rho}_{0}\left[\dfrac{\partial \hat{\psi}_{\bm{\theta}}}{\partial \bm{\varepsilon}^e_I}\left(\bm{s}_{I,0}\right)
    +\bar{\rho}_{0}\mathbf{e}_v\dfrac{\partial \hat{\psi}_{\bm{\theta}}}{\partial \rho}\left(\bm{s}_{I,0}\right)\right]
    -  \bar{\bm{\sigma}}_{I,0}=\boldsymbol{0}, \qquad
    \bm{s}_{I,0} =
        \begin{pmatrix}
        \bar{\rho}_0 & \bm{\varepsilon}^{e}_{I,0} & \bm{\alpha}_0
        \end{pmatrix}^{\top}.
    \label{eq:init_cond}
\end{equation}

Internal variables are initialised to zero, $\bm{\alpha}_{0}=\boldsymbol{0}$, which does not restrict generality since their definition is non-unique in absence of a specified free-energy function (e.g. \citep{flaschel2025convex,masi2024neural}). When microstructural descriptors are available, physically motivated initial values (or learned mappings) can be used instead, e.g. via the multiscale approach in \citep{masi2022multiscale}.
To solve the root-finding problem~\eqref{eq:init_cond}, $\smash{\bm{\varepsilon}^e_{I,0}}$ is treated as an auxiliary learning parameter, which, initialised at zero, is updated by back-propagating the residual $\smash{\bm{\mathcal{R}}}$. This strategy yields smoother optimisation dynamics of the training process and lower computational cost than directly finding the root (at every epoch of the training process).

\subsubsection{Training}
The training process considers a collection of measurement data obtained from different loading protocols indexed by $\ell=1, \ldots, n_{\text P}$. For each protocol $\ell$, stress--strain pairs $\smash{\{ \bar{\bm{\sigma}}_{k}^{\psup},\bar{\bm{\varepsilon}}^{\psup}_{k}\}}$, or equivalently their objective coordinates $\smash{\{ \bar{\bm{\sigma}}_{I,k}^{\psup},\bar{\bm{\varepsilon}}^{\psup}_{I,k}\}}$, are recorded at the discrete physical instants $k=0,\ldots, \smash{n_{\text T}^{\psup}}$, where for notational simplicity $\smash{n_{\text T}^{\psup}}\equiv\smash{n_{\text T}}$ is the number of time samples. Each protocol is also
associated with an initial density $\smash{\bar{\rho}_{0}^{\psup}}$. The set of measurement data is concisely denoted by 
\begin{equation*}
    \bm{\mathcal{D}}= \Bigl\{\bar{\bm{\sigma}}_{I,k}^{\psup},\bar{\bm{\varepsilon}}_{I,k}^{\psup}, \bar{\rho}_0^{\psup} \Bigr\}^{\ell=1,\,\ldots,\,n_{\text P}}_{k=0,\,\ldots,\,n_{\text T}}. 
\end{equation*}

Training consists of identifying network parameters $\bm{\theta}^{\star}$ and the initial value of the elastic strain that minimise the error in the stress predictions and the residual for the initial conditions, namely 
\begin{equation}
    \bm{\theta}^\star = \arg\min_{\bm{\theta}} \left[\frac{1}{n_{\text P}n_{\text T}}\sum_{\ell=1}^{n_{\text P}}\sum_{k=1}^{n_{\text T}}\left\|{\bm{\sigma}}_I(\bm{s}_{I,k}^{\psup}; \bm{\theta})-\bar{\bm{\sigma}}^{\psup}_{I,k}\right\|_2^{2} + \frac{1}{n_{\text P}}\sum_{\ell=1}^{n_{\text P}}\left\|\bm{\mathcal{R}}(\bm{s}_{I,0}^{\psup}, \bar{\bm{\sigma}}_{I,0}^{\psup}; \bm{\theta})\right\|_2^{2}\right],
    \label{eq:loss}
\end{equation}
where $\|\cdot\|_2$ denotes the Euclidean norm applied to the stress coordinates. For the sake of completeness, the stress errors and initial-condition residuals in Eq.~\eqref{eq:loss} are evaluated using normalised quantities, with normalisation parameters computed from the training data. The stress predictions are given by Eq.~\eqref{eq:stress_discrete}, i.e.,
\begin{equation}
    {\bm{\sigma}}_I (\bm{s}_{I,k}^{\psup}; \bm{\theta})= \rho_{k}^{\psup}\left[\dfrac{\partial \hat{\psi}_{\bm{\theta}}}{\partial \bm{\varepsilon}^e_I}(\bm{s}_{I,k}^{\psup}) +\rho_{k}^{\psup}\mathbf{e}_v\dfrac{\partial \hat{\psi}_{\bm{\theta}}}{\partial \rho}(\bm{s}_{I,k}^{\psup})  \right],
    \label{eq:sigma_func}
\end{equation}
and they rely on the solution of the initial value problem \eqref{eq:ivp} to predict the state-space vector $\bm{s}_{I,k}^{\psup}$. For each protocol $\ell$, the chosen time integration scheme produces a discrete trajectory $\smash{\{\bm{s}_{I,n}\}_{n=0}^{N}}$, which is mapped to the physical sampling instants $k=1,\ldots,\smash{n_{\text{T}}}$ by direct matching or interpolation. The distinction between the sampled physical time $k$ and the numerical time steps $n$ (cf.~Section~\ref{subsubsec:solveivp}) allows one to operate with arbitrary sampling frequencies of the data while controlling the accuracy of the time integration through the choice of the numerical time step $h$ and, when needed, substepping between measurement instants.

The minimisation problem~\eqref{eq:loss} is solved by means of the gradient-descent optimisation algorithm Adam in the Python library PyTorch \citep{paszke2017automatic}. Time integration is performed with TorchDiffEq \citep{torchdiffeq} and gradients are evaluated at the converged solutions of the constitutive updates. 

The pseudocode for training is sketched in Algorithm~\ref{alg:training} where the explicit notation for the loading protocols is omitted for conciseness. All loading protocols are evaluated in a single batched call by stacking the initial state vector $\smash{\{\bm{s}_{I,0}^{\psup}\}_{\ell=1}^{n_{\text P}}}$ and integrating all trajectories in parallel. During training, displacement control is assumed and the corresponding strain-rate coordinates $\dot{\bm{\varepsilon}}_I$ are computed using finite differences of the strain coordinates at physical times.

\begin{algorithm}[h]
\caption{Pseudocode of the training algorithm using a set of stress--strain measurement data at time instances $k=1,\ldots,n_{\text T}$ and the initial density. During training, displacement control is assumed and the prescribed strain-rate coordinates are obtained from the measured strain-coordinate histories, for instance by finite differences or by differentiating an interpolant. For the sake of conciseness, the pseudocode considers a single batch/protocol.}
\label{alg:training}
\begin{algorithmic}[1]
\small         

\Require Measurement data ${\bm{\mathcal{D}}= \bigl\{\bar{\bm{\sigma}}_{I,k},\bar{\bm{\varepsilon}}_{I,k}, \bar{\rho}_0 \bigr\}^{n_{\text T}}_{k=0}}$;
\Statex \qquad \, Initial value problem: time integration scheme $\mathtt{solve\_ivp}$ with time step $h$; 
\Statex  \qquad\, Neural networks and initial parameters $\bm{\theta}$: free-energy $\hat{\psi}_{\bm\theta}$ and transport operator $\hat{\;\bm{\mathbb L}}_{\bm\theta}$;%
\Statex \qquad\, Initial conditions: elastic-strain seeds
         $\smash{\bm{\varepsilon}^{e}_{I,0}\gets\bm{0}}$ and internal variables $\smash{\bm{\alpha}_0 \gets \bm 0}$;
\Statex \qquad \, Strain-rate coordinates: \(\dot{\bar{\bm\varepsilon}}_{I}(t)\)
    from \(\{\bar{\bm\varepsilon}_{I,k}\}_{k=0}^{n_T}\)
\Statex \qquad \, Gradient-descent optimiser $\mathtt{optimiser}\,(\{\bm{\theta},\bm{\varepsilon}^{ e}_{I,0}\})$.

\While{\textbf{training}}
  \State Initialise $\mathcal L_{\sigma}\gets0$;
    \State Build initial state and stress: $\bm{s}_{I,0}\gets
           (\bar{\rho}_0,\bm{\varepsilon}^{e}_{I,0},\bm{\alpha}_0)$; $\bm{\sigma}_{I,0}\gets
             {\bm{\sigma}}_I (\smash{\bm{s}_{I,0}}; \bm{\theta}) $

    \State Compute residual loss $\smash{\bm{\mathcal R}\gets \|\bm{\sigma}_{0} -  \smash{\bar{\bm{\sigma}}_{0}}\|_2^2}$
   \Comment{Eq.~\eqref{eq:init_cond}}

\State Solve the initial value problem:
    \(\{\bm s_{I,n}\}_{n=0}^{N}
    \gets
    \mathtt{solve\_ivp}\,
    (\bm s_{I,0},
    \dot{\bar{\bm\varepsilon}}_{I}(t),
    h;\bm\theta)\)
    \Comment{Eq.~\eqref{eq:ivp}}

\State Evaluate the state trajectory at the measurement times:
    \(\bm s_{I,k}\gets \bm s_I(t_k)\)
    
  \For{\(k=1\) \textbf{to} \(n_T\)}
        \State Predict stress:
        \(\bm\sigma_I(\bm s_{I,k};\bm\theta)\)
        \Comment{Eq.~\eqref{eq:sigma_func}}

        \State Accumulate loss \(\mathcal L_{\sigma}
        \gets
        \mathcal L_{\sigma}
        +
        \|\bm\sigma_I(\bm s_{I,k};\bm\theta)
        -
        \bar{\bm\sigma}_{I,k}\|_2^2\)
    \EndFor

  \State $\mathcal L_{\text {tot}} \gets \frac{1}{n_{\text T}}\mathcal L_{\sigma}+ \bm{\mathcal R}$
  \State Optimiser step:  $(\bm{\theta},\bm{\varepsilon}^{e}_{I,0})\gets
         \mathtt{optimiser}\bigl( \{\bm{\theta}, \bm{\varepsilon}^{e}_{I,0}\}; \mathcal L_{\text{tot}}\bigr)$
  \EndWhile
\State \Return trained network parameters $\bm{\theta}^{\star}$
\end{algorithmic}
\end{algorithm}

\subsubsection{Automated selection of internal variables}
\label{subsubsec:internal_var}

Depending on the information collected in the training measurements, an accurate prediction of the evolution of the stress may require one or more internal variables in addition to the elastic strain and the mass density. The number of internal variables is chosen via a discrete search over the value of $n_{\bm\alpha}=0,1,\ldots$ by selecting the smallest dimension of $\bm{\alpha}$ that guarantees a low (validation) error for a given architecture $(\hat{\psi}_{\bm{\theta}},\hat{\;\bm{\mathbb{L}}}_{\bm{\theta}})$, see \citep{masi2023evolution}. For each candidate $n_{\bm\alpha}$, the model is trained by minimising the right-hand side of Eq.~\eqref{eq:loss} and assessed on held-out protocols using the error in the stress predictions. The selected value $n_{\bm\alpha}^\star$ is the smallest one such that increasing $n_{\bm\alpha}$ does not yield a significant improvement in the performance to balance accuracy and generalisation. Note that the internal variables are not preassigned physical quantities, with their nature and values being learned end-to-end, but thermodynamic admissibility holds for any value $n_{\bm{\alpha}}$, by construction. As a consequence, they provide a latent, energetically consistent parametrisation of the history- and path-dependent material behaviour (cf.~Section~\ref{subsubsec:init_cond}).

\subsubsection{Inference}
After training, the two neural networks $\hat{\psi}_{\bm{\theta}^{\star}}$ and $\hat{\;\bm{\mathbb{L}}}_{\bm{\theta}^{\star}}$ are employed to infer the material response corresponding to new, unobserved loading protocols. The pseudocode for inference is sketched in Algorithm~\ref{alg:inference}. At prediction time, the initial elastic-strain coordinates are obtained by solving Eq.~\eqref{eq:init_cond} with the trained parameters fixed. For strain-controlled protocols, the prescribed strain-rate coordinates are computed from the imposed strain history. For stress-controlled or mixed-control protocols, the constitutive update is augmented with the corresponding algebraic stress constraints and solved by Newton or quasi-Newton iterations, with a prescribed relative tolerance of $10^{-6}$.

\subsubsection{Identifiability and interpretation of learned representations}
The inverse problem of identifying the network parameters $\bm{\theta}$ that minimise Eq.~\eqref{eq:loss} is severely under-determined. For any finite collection of stress trajectories $\bar{\bm{\sigma}}_I \in \bm{\mathcal{D}}$, there exists an infinite family of network parameters and material states that reproduce these data,
\begin{equation}
    \bm{\mathcal{M}}^{\star} = \{( \bm{\theta}^{\star},\bm{s}_I) \mid  {\bm{\sigma}}_I(\bm{s}_I; \bm{\theta}^{\star}) = \bar{\bm{\sigma}}_I\}.
\end{equation}
Among these uncountably many equivalent representations, the optimisation problem~\eqref{eq:loss} is restricted to the physically admissible set $\bm{\mathcal{M}}^{\star}_{\text{adm}} \subseteq \bm{\mathcal{M}}^{\star}$ that enforces by construction the conservation of mass and energy, the dissipation inequality, and the principle of objectivity.

Since the optimisation in~\eqref{eq:loss} is restricted to $\bm{\mathcal M}^{\star}_{\mathrm{adm}}$, the trained network delivers a thermodynamically consistent mapping $\bm{s}_I\;\mapsto\;\bm{\sigma}_I\bigl(\bm{s}_I;\bm{\theta}^{\star}\bigr)$ for any admissible loading path. The pair $\smash{(\hat{\psi}_{\bm{\theta}^{\star}},\hat{\;\bm{\mathbb L}}_{\bm{\theta}^{\star}})}$ should therefore be interpreted as one admissible constitutive representation, rather than the unique material law. The same consideration applies to the identified elastic strain and internal variables, whose values additionally depend on the specified initial conditions (cf.\ Sect.~\ref{subsubsec:init_cond}). The initial elastic-strain coordinates follow the root-finding problem~\eqref{eq:init_cond} which is well-posed for a given $\bm{\theta}^{\star}$ under the conditions considered, while all internal variables are initialised to zero. This choice deliberately avoids the phenomenological assumptions often adopted in conventional constitutive modelling and identifies a thermodynamically admissible state space when no microstructural information is available.

Note that the framework can be extended to identify internal variables from the internal structure of materials. When richer measurements are available -- for instance, topological descriptors or internal degrees of freedom of the material microstructure -- the admissible set $\bm{\mathcal M}^{\star}_{\mathrm{adm}}$ can be further restricted by linking the internal variables to those data, following the methodology proposed in~\citep{masi2022multiscale,masi2023evolution,masi_alert}.

\begin{algorithm}[h]
\caption{Pseudocode of the inference algorithm relying on the trained network parameters $\bm{\theta}^{\star}$, the initial values for stress and density, and the strain time-history or force control.}
\label{alg:inference}
\begin{algorithmic}[1]
\small

\Require Measured strain histories $\{\bar{\bm{\varepsilon}}_{I,k}\}$ (or stress control $\{\bar{\bm{\sigma}}_{I,k}\}$),
        initial stresses ${\bm{\sigma}}_{I,0}$,
        initial density ${\rho}_0$;
\Statex \qquad \, Prescribed loading/control;
\Statex \qquad \, Root-finding solver $\mathtt{root\_solver}$, initial-value-problem routine and method $\mathtt{solve\_ivp}$ and time step $h$;
\Statex  \qquad\, Trained parameters $\bm\theta^{\star}$ (neural networks $\hat{\psi}_{\bm\theta^{\star}},\hat{\;\bm{\mathbb L}}_{\bm\theta^{\star}}$).
    
  \State Initialise $\bm{s}_{I,0} \gets  ({\rho}_0,\bm{0},\bm{0})$
  \State  Determine the initial elastic-strain coordinates by solving $\bm{s}_{0}\gets
         \mathtt{root\_solver}\left( \bm{\mathcal{R}}\left(\bm{s}_{I,O}, {\bm{\sigma}}_{I,0}; \bm{\theta}^{\star}\right)\right)$
         \Comment{Eq.~\eqref{eq:init_cond}}

 \If{strain control}
\State Compute prescribed strain-rate coordinates
\(\dot{\bar{\bm\varepsilon}}_I(t)\) from the imposed strain history
  \State Solve the initial value problem: \(\{\bm s_{I,n}\}_{n=0}^{N} \gets \mathtt{solve\_ivp}\, (\bm s_{I,0}, \dot{\bm\varepsilon}_I(t), h;\bm\theta^\star)\)
 \Comment{Eq.~\eqref{eq:ivp}}
    \For{$n = 0$ \textbf{to} $N$}
        \State Predict stress: $\bm{\sigma}_{I,n} \gets {\bm{\sigma}}_I\bigl(\bm{s}_{I,n};\bm\theta^{\star}\bigr)$
        \Comment{Eq.~\eqref{eq:sigma_func}}
    
    \EndFor
    
\ElsIf{stress control}
\State Solve Eq.~\eqref{eq:ivp} with the control constraint for the unknown strain-rate coordinates
\State Store the resulting trajectories:
\( \{\bm s_{I,n}\}_{n=0}^{N}\) and \( \{\bm \sigma_{I,n}\}_{n=0}^{N}\)
\EndIf

\State \Return stress-coordinate history $\{\bm{\sigma}_{I,n}\}_{n=0}^{N}$ and state trajectory $\{\bm{s}_{I,n}\}_{n=0}^{N}$
\end{algorithmic}
\end{algorithm}

\section{Synthetic benchmarks: identification of constitutive equations}
\label{sec:benchmarks}

The performance and accuracy of the proposed methodology are investigated and quantified relying on synthetic data generated from analytical constitutive laws. These benchmarks are designed as a validation ladder of increasing difficulty, i.e., elasto- and hypo-plasticity, including rate and hardening effects, and non-associative plastic flows. Having exact knowledge of the ground truth, i.e., the constitutive model that underlies the data, these tests allow a strict and quantitative assessment of the proposed methodology. In particular, they test the ability to identify consistent constitutive equations from data, infer unobserved material responses, and identify quantities that are not provided during training, including the free energy, the dissipation rate, and internal variables. To highlight the benefit of enforcing non-equilibrium thermodynamics as hard constraints via generalised transport equations, no prior information about internal variables is used following the automatic selection procedure detailed in Section~\ref{subsubsec:internal_var}.

In the considered synthetic datasets, only stress--strain pairs are sampled. Here, density is assumed constant and the identification focuses on the inelastic evolution. The initial value problem Eq.~\eqref{eq:ivp} is integrated over a unit pseudo-time interval using an explicit scheme (explicit Euler), with step size $h=4.0 \cdot 10^{-3}$, i.e., 250 steps per protocol. 

For completeness, Appendix~\ref{app:hard_soft_dissipation} provides a comparison of the present hard-constrained framework with learning formulations using soft constraints for the dissipation inequality.

\subsection{One-dimensional benchmarks}

The first set of benchmarks involves the following one-dimensional material laws: elasto-plasticity (EP), hypo-plasticity (HP), elasto-visco-plasticity with Perzyna-type viscosity (EVP), and elasto-plasticity with isotropic hardening (EPH).

\begin{figure}[ht]
\centering
\includegraphics[width=0.9\textwidth]{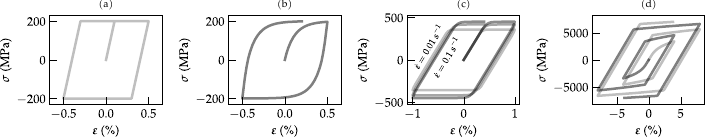}
\caption{Loading protocol and stress data for the one-dimensional benchmarks used for training. From left to right: (a) elasto-plastic (EP), (b) hypo-plastic (HP), (c) elasto-visco-plastic (EVP), and (d) elasto-plastic with isotropic hardening (EPH) material.}
\label{fig:1d_materials}
\end{figure}

\subsubsection{Elasto-plasticity}
The first example concerns an idealised elastic, perfectly plastic material. The stress depends only on the elastic strain, $\sigma=E\,\varepsilon^{e}$, with yield condition $|\sigma|\le \sigma_y$. The material has elastic modulus $E=200$~GPa and yield stress $\sigma_y=200$~MPa.

\paragraph{Data and training}
The training set consists of one protocol with sequential loading--unloading--reloading at constant strain-rate amplitude $|\dot{\varepsilon}|=0.01$~s$^{-1}$, see Figure~\ref{fig:1d_materials}. The free-energy network $\smash{\hat{\psi}_{\bm\theta}}$ has two hidden layers with 64 units and exponential-linear-unit (ELU) activation; the transport-operator network $\smash{\hat{\;\bm{\mathbb L}}_{\bm\theta}}$ has three hidden layers with 64 units and leaky-rectified linear unit (LReLU) activation. The training loss without considering internal variables converges after $\approx 3000$ epochs, with a mean absolute error in the stress predictions below $0.1\%$. 

\paragraph{Results}

\begin{figure}[ht!]
\centering
\includegraphics[width=0.9\textwidth]{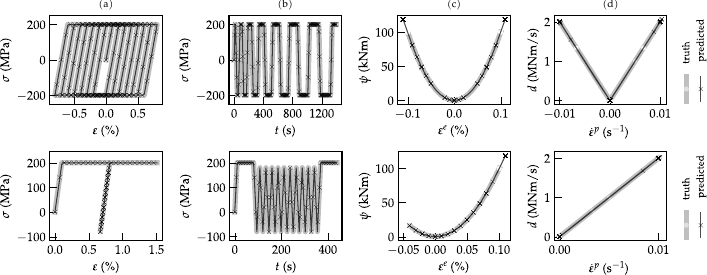}
\caption{Predictions at inference of the response of the elasto-plastic (EP) material under unobserved loading protocols: increasing-amplitude cyclic test (top) and stress cycles of fixed amplitude (bottom). From left to right: (a) stress versus strain ($\sigma$--$\varepsilon$), (b) stress versus time ($\sigma$--$t$), (c) free energy versus elastic strain ($\psi$--$\varepsilon^e$), and (d) dissipation rate versus plastic strain rate ($d$--$\dot{\varepsilon}^p$). The free energy is reported per unit mass.}
\label{fig:EP_inf}
\end{figure}

Figure~\ref{fig:EP_inf} compares the predictions at inference with the ground-truth results for two unobserved loading paths: (\emph{i}) strain cycles of increasing amplitude and (\emph{ii}) stress cycles of fixed amplitude. Figure~\ref{fig:EP_inf}~(a) shows the stress--strain response, (b) the time evolution of the stress, (c) the evolution of the inferred free energy with respect to the identified elastic strain, and (d) the dissipation rate versus the inferred plastic strain rate. The free energy and dissipation rate are reported per unit mass, unless stated otherwise.

Under strain cycling (Fig.~\ref{fig:EP_inf}, top), the neural material model reproduces the onset of yielding at the expected elastic threshold $\sigma_y/E=0.1\%$ and the subsequent perfectly plastic response, including the characteristic stress plateau and the return to the same elastic slope upon unloading. The size of the elastic domain and the plastic flow direction remain consistent across cycles, and the predicted plastic strain rate, free energy and dissipation rate match the reference evolution.

For stress cycles of fixed amplitude with $\max|\sigma|<\sigma_y$ (Fig.~\ref{fig:EP_inf}, bottom), the model correctly predicts a purely reversible elastic response with closed $(\sigma,\varepsilon)$ loops, vanishing plastic strain rates, and zero dissipation, see Figure~\ref{fig:EP_inf}~(c,d). These results show that, even when trained on a single simple protocol and stress-only data, hard thermodynamic constraints enable generalisation of both the macroscopic response and the thermodynamic fields to loading modes not seen during training.

\subsubsection{Hypo-plasticity}
The second example considers a hypo-plastic material model with emerging ratcheting \citep{kolymbas1991outline} governed by the nonlinear stress-rate equation $\dot{\sigma} = E\dot{\varepsilon} + E(\sigma/\sigma_y)^{s} |\dot{\varepsilon}|$, where the elastic modulus and yield stress are kept the same as in the first benchmark and $s=0.5$ expresses the nonlinearity of the hypo-plastic formulation. In contrast to elasto-plasticity, hypo-plasticity does not rely on elastic-plastic strain-rate decomposition. Moreover, the plastic flow rule cannot be cast as the (sub-)differential of a convex dissipation potential (cf. \citep{einav2012unification}). These aspects make this simple case a stringent test for the proposed methodology.
\paragraph{Data and training}
The training set, loading protocol, and network architecture are kept identical to those of the elasto-plastic benchmark, see Figure~\ref{fig:1d_materials}.

\begin{figure}[ht!]
\centering
\includegraphics[width=0.9\textwidth]{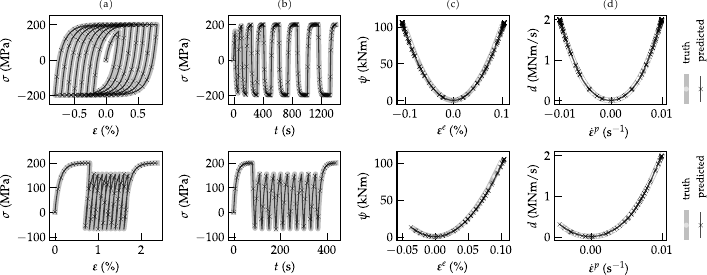}
\caption{Predictions at inference of the response of the hypo-plastic (HP) material under unobserved loading protocols: increasing-amplitude cyclic test (top) and stress cycles with ratcheting (bottom). From left to right: (a) stress versus strain ($\sigma$--$\varepsilon$), (b) stress versus time ($\sigma$--$t$), (c) free energy versus elastic strain ($\psi$--$\varepsilon^e$), and (d) dissipation rate versus plastic strain rate ($d$--$\dot{\varepsilon}^p$).}
\label{fig:HP_inf}
\end{figure}

\paragraph{Results}
Figure~\ref{fig:HP_inf} shows the predictions and the reference results for the hypo-plastic material model for the same two families of unobserved paths as for the case of the elasto-plastic model. The neural model attains high accuracy in terms of the stress and the inferred elastic strain, free energy, plastic strain rate, and dissipation rate.

Under strain cycling of increasing amplitude (Fig.~\ref{fig:HP_inf}, top), the predicted $(\sigma,\varepsilon)$ trajectories show that the tangent stiffness evolves smoothly with the stress level, consistent with the nonlinear rate equation. The dissipation rate remains strictly positive, reflecting the non-recoverable character of the response, and the inferred elastic strain and free energy evolve consistently throughout the cycles.

Under stress cycling (Fig.~\ref{fig:HP_inf}, bottom), the network captures the hallmark of hypo-plasticity, ratcheting, i.e., the progressive, non-hysteretic accumulation of strain with each cycle. The model reproduces the gradual drift in total strain and the open $(\sigma,\varepsilon)$ trajectories associated with ratcheting.

\subsubsection{Rate dependence}
The third benchmark builds upon the elasto-plastic formulation of the first benchmark by further introducing rate dependence through Perzyna's overstress law (see \citep{stathas2022role}). The elastic modulus and yield stress are equal to 100 GPa and 200 MPa, respectively, and are complemented with the viscosity parameter equal to $1$ GPa/s and the rate-sensitivity dimensionless exponent, set equal to $5$.
\paragraph{Data and training}
The training set consists of one protocol with a sequence of loading--unloading--reloading at three different constant strain-rate amplitudes $|\dot{\varepsilon}| = 0.01$, $0.05$, and $0.1\%$ s$^{-1}$, see Figure~\ref{fig:1d_materials}. The neural network architecture is the same as for the previous benchmarks.

\begin{figure}[ht]
\centering
\includegraphics[width=0.9\textwidth]{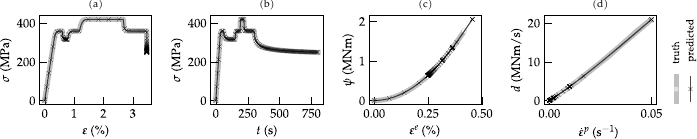}
\caption{Predictions at inference of the response of the elasto-visco-plastic (EVP) material under velocity stepping and subsequent hold test, i.e., $\dot{\varepsilon}=\{0.01,0.002,0.01,0.05,0.0\}$ s$^{-1}$. From left to right: (a) stress versus strain ($\sigma$--$\varepsilon$), (b) stress versus time ($\sigma$--$t$), (c) free energy versus elastic strain ($\psi$--$\varepsilon^e$), and (d) dissipation rate versus plastic strain rate ($d$--$\dot{\varepsilon}^p$).}
\label{fig:EVP_inf}
\end{figure}

\paragraph{Results}
Figure~\ref{fig:EVP_inf} showcases the predictions of the neural model under a velocity-stepping sequence followed by a hold at zero strain rate, none of which appeared in training. The neural model identifies the expected rate sensitivity: larger $|\dot{\varepsilon}|$ yields larger overstress, with stress levels and plastic strain-rate values correctly matching the reference response across all steps. During the hold phase, the model captures the stress relaxation toward the yield value, the decay of the plastic strain rate, and the vanishing of the dissipation rate. The inferred state variable, $\varepsilon^e$, also matches the reference model, during both the rate transitions and the relaxation. The same holds for the free energy. The results indicate that the learned dynamics encode the correct rate-overstress coupling and generalise to relaxation phenomena absent from the training data.

\subsubsection{Elasto-plasticity with hardening}
\label{subsubsec:EPH}
The fourth benchmark considers an elasto-plastic constitutive law with Ramberg--Osgood isotropic hardening \citep{ramberg1943description}, with yield condition $\smash{|\sigma| \leq \sigma_y + C (\varepsilon^p_c)^m}$, where $\varepsilon^p_c$ is the cumulative plastic strain, $C=10$~MPa the hardening modulus, and $m=0.3$ is the hardening exponent. The elastic modulus and yield strength are $E=200$~GPa and $\sigma_y=200$~MPa.

\paragraph{Data and training}
The training set consists of two protocols with sequential loading--unloading  phases at a constant strain-rate amplitude $|\dot{\varepsilon}|=0.002~\mathrm{s}^{-1}$, see Figure~\ref{fig:1d_materials}. The network architecture remains unchanged. An initial run assumes a state space spanned solely by the elastic strain. Under this assumption, the mean absolute percentage error stagnates at $\approx 142\%$ after a few hundred epochs, indicating that a single-variable state is insufficient to represent isotropic hardening (cf. Sect.~\ref{subsubsec:internal_var}). Training is then repeated introducing a scalar internal variable $\alpha$, learned end-to-end without a prescribed interpretation. With this modification, the mean absolute error decreases to $\approx 0.2\%$, confirming that one internal variable is enough to capture the observed material response.

\begin{figure}[th]
\centering
\includegraphics[width=0.9\textwidth]{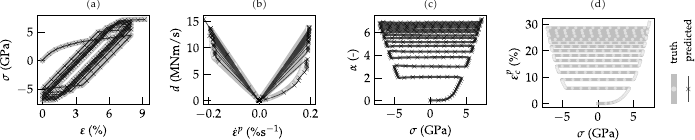}
\caption{Predictions at inference of the response of the elasto-plastic material with isotropic hardening (EPH) under an unobserved cyclic loading protocol with multiple loading and unloading sequences. From left to right: (a) stress versus strain ($\sigma$--$\varepsilon$), (b) dissipation rate versus plastic strain rate ($d$--$\dot{\varepsilon}^p$), (c) automatically identified internal variable $\alpha$ and (d) cumulative plastic strain $\varepsilon^p_c$ versus stress.}
\label{fig:EPH_inf}
\end{figure}

\paragraph{Results}
Figure~\ref{fig:EPH_inf} evaluates the generalisation of the neural model to an unseen multi-cycle path. The model reproduces the defining features of isotropic hardening with the progressive expansion of the yield level with accumulated plastic strain. The dissipation rate versus plastic strain rate ($d$--$\dot{\varepsilon}^p$, panel b) curves agree closely with the reference model and minor discrepancies are limited to a few localised points near load reversals. The automatically identified internal variable $\alpha$ behaves as a scalar hardening measure, nearly monotone during plastic flow and ordering the cycles by the expansion of the elastic domain, correlating strongly with the cumulative plastic strain $\varepsilon^p_c$ (cf. panels c, d).

\begin{figure}[th]
\centering
\includegraphics[width=0.9\textwidth]{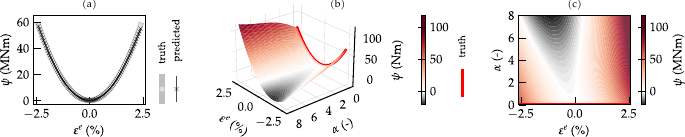}
\caption{Predictions at inference of the response of the elasto-plastic material with isotropic hardening (EPH) in terms of the free energy $\psi$. From left to right: (a) comparison between predictions and ground truth for $\alpha=0$ and zero cumulative plastic strain, respectively, and (b--c) predicted energy as a function of the elastic strain and the identified internal variable.}
\label{fig:EPH2_inf}
\end{figure}

Figure~\ref{fig:EPH2_inf} presents the inferred free energy. At $\alpha=\varepsilon^p_c = 0$, the energy closely matches the reference elastic energy (panel a). The learned surface $\psi(\varepsilon^{e},\alpha)$ increases with both $|\varepsilon^{e}|$ and $\alpha$, resulting in stress values consistent with the hardening observed in Fig.~\ref{fig:EPH_inf}. A small asymmetry in the free energy with respect to the elastic strain is observed (panel c), but the predictions at inference remain physically consistent, indicating that the model has discovered an interpretable hardening variable and a consistent energy landscape from stress--strain data.

\subsection{Three-dimensional non-associative benchmark}
\label{subsec:3D-DP}
The last benchmark showcases a three-dimensional, incrementally nonlinear, elasto-plastic, non-associative Drucker--Prager material with constant density (cf. \citep{einav2012unification}). The material parameters are: bulk modulus $K=46.7$~MPa, shear modulus $G=40$~MPa, friction and dilatancy coefficients equal to $1.0$ and $0.5$, respectively, and seismic parameter equal to $100$. The neural constitutive equations are formulated using the invariant coordinates introduced in Section~\ref{subsec:hard_constraints}, $\bm\sigma_I=(p\quad q)^\top$,
$\bm\varepsilon_I=(\varepsilon_v \quad \varepsilon_s)^\top$, with compression taken as positive. 
\begin{figure}[ht]
\centering
\includegraphics[width=0.9\textwidth]{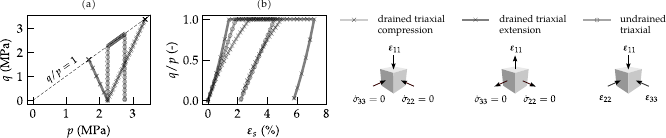}
\caption{Loading protocols for the training process in terms of (a) deviatoric versus volumetric stress ($q$--$p$) and (b) friction versus deviatoric strain ($q/p$--$\varepsilon_s$). The measurement data include: drained triaxial compression/extension ($\varepsilon_{11}$ prescribed, $\dot{\sigma}_{22} = \dot{\sigma}_{33} = 0$) and (isochoric) undrained triaxial compression ($\varepsilon_{11}$ prescribed, $\varepsilon_{22}=\varepsilon_{33} = -\frac{\varepsilon_{11}}{2}$) at different initial confining pressures $p$.}
\label{fig:3d_materials}
\end{figure}

\paragraph{Data and training}
Training data are generated by single loading--unloading  paths, as in conventional laboratory testing. From an initially unstressed state ($\bm\sigma_I = \boldsymbol{0}$), isotropic compression first creates admissible states at several confining pressures ($0.5 \le p \le 2$ MPa, $q=0$ MPa). These states then serve as initial conditions for drained triaxial compression and extension tests and undrained triaxial compression tests, depicted in Figure~\ref{fig:3d_materials}, with strain rates of order $0.3$--$1\times10^{-3}$ s$^{-1}$.

The free-energy network $\hat{\psi}_{\bm{\theta}}$ uses three hidden layers with 64 nodes and ELU activation and the transport-operator network $\hat{\;\bm{\mathbb{L}}}_{\bm{\theta}}$ uses four hidden layers with 64 nodes and Softplus activation. The material state is assumed to be spanned by the elastic strain only. With this choice, the training loss converges in approximately $5000$ epochs to a mean absolute error in the stress predictions below $0.1\%$, indicating that no additional internal variables are required.

\paragraph{Results}
Predictions at inference are tested for an unseen mixed-control, cyclic path, i.e., simple shear, where $\varepsilon_{12}$ is prescribed and normal stresses are kept constant from an initial confining pressure $p=2$ MPa. The stress-control condition (constant $p$) is enforced at every step by Newton--Raphson iterations with the neural material model.

\begin{figure}[ht]
\centering
\includegraphics[width=0.9\textwidth]{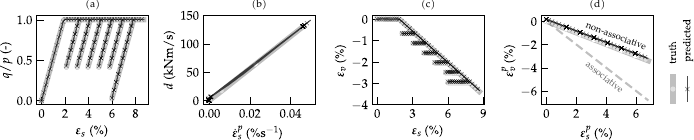}
\caption{Predictions at inference for an unobserved protocol (simple shear, i.e., $\varepsilon_{12}$ prescribed, $\dot{\sigma}_{11}=\dot{\sigma}_{22}=\dot{\sigma}_{33}=0$) with multiple loading--unloading  sequences. From left to right: (a) friction versus shear strain ($q/p$--$\varepsilon_s$), (b) dissipation rate versus shear plastic strain rate ($d$--$\dot{\varepsilon}^p_s$), (c) volumetric versus shear strain ($\varepsilon_v$--$\varepsilon_s$), and (d) volumetric versus shear plastic strain ($\varepsilon_v^p$--$\varepsilon_s^p$).}
\label{fig:DP_inf}
\end{figure}

Figure~\ref{fig:DP_inf} shows the predictions in terms of the shear stress, dissipation rate, and volumetric strain. The friction response $q/p$ versus the shear strain $\varepsilon_s$ matches the reference results over the repeated loading--unloading  cycles (panel a). The dissipation versus plastic-shear rate relationship $d$--$\dot{\varepsilon}^p_s$ (panel b) is strictly non-negative and scales with $|\dot{\varepsilon}^p_s|$ as expected for a rate-independent plastic flow with only limited discrepancies with respect to the reference model near load reversals. The predicted volumetric-shear trajectories (panels c--d) demonstrate the ability of the framework to retrieve the inherent non-associative behaviour. The plastic volumetric-shear strain trajectory displays a slope consistent with the nominal dilatancy coefficient ($0.5$) in contrast with a purely associative flow which is included for reference. Non-associativity is recovered naturally from the data and the identified transport operator provides a general evolution law with thermodynamic admissibility guaranteed by construction.

\begin{figure}[ht]
\centering
\includegraphics[width=0.765\textwidth]{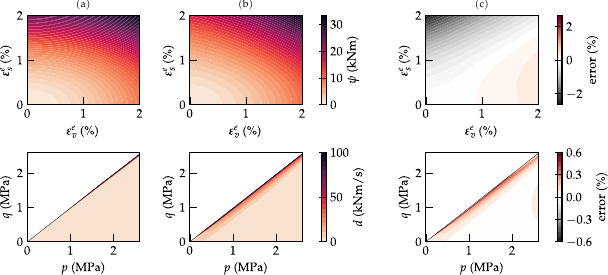}
\caption{Ground-truth and learned free energy function (top) and dissipation rate (bottom). From left to right: (a) true and (b) learned free energy (top) and dissipation rate (bottom); (c) percentage relative error for the free energy (top), $(\psi_{\bm{\theta}}-\bar{\psi})/\bar{\psi}_{\textrm{max}} \,\%$, and dissipation rate (bottom), $(d_{\bm{\theta}}-\bar{d})/\bar{d}_{\textrm{max}} \,\%$, where $\psi_{\theta}$ and $d_{\bm{\theta}}$ are the predictions, $\bar{\psi}$ and $\bar{d}$ are the analytical values, and the subscript $\text{max}$ denotes the maximum value.}
\label{fig:DP_inf2}
\end{figure}

Figure~\ref{fig:DP_inf2} compares the identified thermodynamic fields with their analytical counterparts, namely the free energy (top) and the dissipation rate (bottom). The inferred free energy $\psi$ is convex in the selected elastic-strain invariant coordinates by construction. For elastic-strain magnitudes up to $\approx 1\%$, it coincides with the reference values. For larger magnitudes, small deviations appear (relative error within $\approx 2\%$), consistent with a slight offset in the effective elastic moduli relative to the nominal $(K,G)$. The learned and reference dissipation rate are presented in Figure~\ref{fig:DP_inf2} (bottom) considering a field of admissible stress states ($p,q$). 
The dissipation rate is everywhere non-negative and its magnitude increases as $q/p$ approaches 1 (yield locus). Quantitatively, the percentage relative error stays below $\approx\!0.6\%$ with largest deviations confined near the yield locus where the inferred dissipation is slightly higher than the reference. 

Finally, it is instructive to examine the structure of the learned force--flux map, i.e., $ \bm z_I = \mathbf{{L}}_{\bm\theta}(\bm{\mathcal Y}_I,\bm s_I)\bm{\mathcal Y_I}$, with $\bm{\mathcal Y}_I \equiv \bm{\sigma}^e_I$, $\bm z_I \equiv \dot{\bm{\varepsilon}}^p_I$, and $\mathbf L_{\bm\theta}$ denoting the matrix representation of the learned transport operator $\smash{\hat{\;\bm{\mathbb L}}_{\bm\theta}}$. The resulting force--flux Jacobian is $\smash{{\bm{{J}}}_{\bm{\theta}}= \partial \bm z_I/\partial \bm{\mathcal Y}_I = {\mathbf{{L}}}_{\bm\theta} + \bm{C}_{\bm\theta}}$ and contains both the direct transport contribution
$\mathbf L_{\bm\theta}$ and the additional contribution $\bm C_{\bm\theta}\equiv (\partial{\mathbf{{L}}}_{\bm\theta}/ \partial \bm{\mathcal{Y}}) \bm{\mathcal{Y}}$ induced by the force dependence of the transport operator. 
Since force--flux relations generated by scalar dissipation potentials have a symmetric Jacobian, the skew-symmetric part of $\bm J_{\bm\theta}$ provides a local measure of the non-potential behaviour and highlights the additional force--flux couplings made available by generalised transport equations (cf. Section~\ref{subsubsec:potential}). To quantify this, the following metrics are introduced:
\begin{equation}
\varkappa_{J} =\frac{
\left\|\operatorname{Skw}({\bm{{J}}})\right\|_F
}{ \left\|{\bm{{J}}}\right\|_F
}, \quad \varkappa_{L} =\varkappa_J\frac{
\|\operatorname{Skw}(\mathbf{L}_{\bm{\theta}})\|_F
}{ \|\operatorname{Skw}(\mathbf{L}_{\bm{\theta}})\|_F + \left\|\operatorname{Skw}(\bm{C}_{\bm{\theta}})\right\|_F},
\quad \varkappa_{C} =\varkappa_J \frac{
\|\operatorname{Skw}(\bm{C}_{\bm{\theta}})\|_F
}{ \left\|\operatorname{Skw}(\mathbf{L}_{\bm{\theta}})\right\|_F + \left\|\operatorname{Skw}(\bm{C}_{\bm{\theta}})\right\|_F},
\label{eq:jacobian_asymmetry}
\end{equation}
where $\|\cdot\|_F$ is the Frobenius norm. The quantity $\varkappa_J$ measures the relative asymmetry of the local force--flux Jacobian, while $\varkappa_{L}$ and $\varkappa_{C}$ quantify the contributions associated with the skew part of the transport operator and with the skew part induced by its force-dependence, respectively.

\begin{figure}[h!]
\centering
\includegraphics[width=0.729\textwidth]{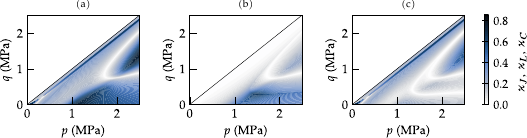}
\caption{Structure of the learned force--flux map within the admissible domain \(q/p<1\). From left to right: (a) skew-symmetric component \(\varkappa_J\) of the force--flux Jacobian $\bm{J}_{\bm{\theta}}$, (b) skew component \(\varkappa_L\) associated with the explicit skew contribution of the transport operator, and (c) skew component \(\varkappa_C\) associated with its force-dependence, cf. Eq.~\eqref{eq:jacobian_asymmetry}.}
\label{fig:rJ}
\end{figure}

Figure~\ref{fig:rJ} displays the structure of the learned force--flux map over the admissible stress domain $q/p<1$. The map of $\varkappa_J$ shows that the learned transport representation has a spatially heterogeneous structure in the $(p,q)$ plane. Regions where the Jacobian is nearly symmetric, $\varkappa_J \approx 0$, coexist with regions where its skew part is substantial, e.g. near the $q/p=1$ domain and in parts of the high-confinement region. The components $\varkappa_L$ and $\varkappa_C$ clarify how this structure is expressed by the learned transport representation. Most of the non-potential coupling is associated with the force-dependence term $\bm C_{\bm\theta}$, whereas the explicit skew contribution of $\mathbf L_{\bm\theta}$ remains comparatively small. This behaviour is consistent with the non-associative character of the present material: the identified evolution law mainly uses the force dependence of the transport operator to encode force--flux couplings beyond those generated by scalar potentials.

\section{Application to granular materials: discovery of constitutive equations}
\label{sec:discovery}
Granular media are excellent candidates to test the performance of the present learning framework, as they are heterogeneous materials that exhibit a rich, complex, and history-dependent behaviour. Classical quasi-static material models rely on heuristic plasticity and critical-state theory (e.g. state-dependent friction, dilatancy relations \citep{kolymbas2000misery,schofield1968critical,taiebat2008sanisand,alaei2021hydrodynamic,tafili2024hypoplastic}), yet they remain hindered by the phenomenological choices of the state variables and underlying constitutive relations, which are often tuned by human trial-and-error adjustments. Despite the progress enabled by in operando testing \citep{vlahinic2014towards,kawamoto2018all,tengattini2023micromechanically}, no universally accepted closed-form law exists for granular materials, particularly across transitional regimes (e.g. quasi-static to dynamic) and under cyclic, non-proportional loading paths. 

Here, the learning framework is applied to the discovery of constitutive equations for a dry, cohesionless granular material under virtual testing conditions that mimic laboratory settings. Laboratory data typically reflect a combination of (\emph{i}) aleatoric (statistical) uncertainty from the material's inherent heterogeneity, (\emph{ii}) epistemic uncertainty arising from incomplete knowledge of the material state and boundary conditions, and (\emph{iii}) measurement errors. These challenges are often exacerbated by a sparse coverage of the state space (few tested protocols and initial conditions) and limited temporal resolution. In the present work, attention is primarily focused on aleatoric and epistemic sources of uncertainty.

Physical tests are replaced by in silico experiments using the Discrete Element Method (DEM, \citep{cundall1979discrete}) as a controlled virtual simulator to generate high-fidelity numerical analogues of granular material responses, rather than as a replacement for laboratory experiments. DEM explicitly represents the micro-mechanical and geometrical structure of the granular assembly, so that the macroscopic response emerges from prescribed particle properties, contact laws, and particle interactions rather than from an imposed macro-mechanical constitutive law. Aleatoric variability arises from spatial heterogeneity in material properties and microstructural topology. Epistemic uncertainty is associated with the limited observability of the internal material state, such as fabric and force chains, in conventional laboratory settings. Although DEM provides access to such grain-scale information, the present study deliberately restricts the measurement data to macroscopic observables readily available in laboratory tests, namely stress--strain pairs and the initial mass density. The unobserved grain-scale information is therefore treated as epistemic. 

\subsection{Discrete element method and in silico experiments}
\label{subsec:DEM_exp}
The reference granular material is modelled as an assembly of spherical discrete-element particles. The particles interact with each other through frictional contact laws and their motion is governed by Newton's laws of motion. The numerical experiments are conducted using the open-source software YADE-Open DEM \citep{kozicki2009yade}, where the equations of motion are integrated in time explicitly using a central finite difference algorithm.

The analogue material is a purely frictional granular medium. The adopted particle properties define an effective DEM analogue of a dry, cohesionless granular assembly under quasi-static loading. Each grain is assigned a diameter $d$, a mass density $\rho_s=2600$ kg/m$^3$, and an effective elastic behaviour defined by a Young's modulus $E=60$ MPa and a Poisson's ratio $\nu=0.5$. Grains interact through Coulomb friction interfaces with friction coefficient $\mu=0.5$. The grain-size distribution is sampled from a uniform law $d\sim d_{50}\mathcal{U}(0.5,1.5)$, with median diameter $d_{50}=10$~\textmu m. 

The present application is restricted to quasi-static deformation regimes, by keeping the dimensionless inertial number $I=\smash{\dot{\gamma}d_{50} \sqrt{\rho/p}}$ below a value of $10^{-4}$ \citep{midi2004dense}, where $\dot{\gamma}$ is the shearing rate, $d_{50}$ the median particle diameter, $\rho$ the mass density, and $p$ the confining pressure. To limit boundary effects and satisfy the Hill--Mandel condition, the simulated granular packings employ periodic boundary conditions \citep{o2011particulate}.

The DEM simulations are treated here as material-point virtual tests expressed in engineering stress and strain measures, as commonly done in triaxial testing of geomaterials. The learned model is therefore interpreted within the infinitesimal-strain setting of Section~\ref{sec:theory}. The approach can be extended to finite-deformation kinematics by replacing the infinitesimal strain measures and state laws with their finite-deformation counterparts, see e.g. \cite{holthusen2025generalized}.

Let $V$ be a material element, under periodic boundary conditions and quasi-static conditions. The volume-averaged Cauchy stress is given by the Love--Weber expression \citep{love1892,weber1966recherches}:
\begin{equation}
  \bm{\sigma}_V \equiv \smash{\frac{1}{|V|}\int_V \bm{\sigma}(\mathbf{x})\, \text{d}\mathbf{x}}= \frac{1}{|V|} \sum_{c} \bm{f}_c \otimes \bm{l}_c,
\end{equation}
where $|V|$ is the volume, $\bm{f}_c$ is the contact force, and $\bm{l}_c$ is the branch vector, with summation over all contacts $c$ in the assembly. The average strain rate and strain are directly obtained from the boundary deformation of the periodic cell. Under periodic boundary conditions, the dissipation inequality~\eqref{eq:clausius_} can be reformulated in terms of average quantities
\begin{equation}
    d_{V} = \bm{\sigma}_V\! : \! \dot{\bm{\varepsilon}}_V - \rho_V \dot{\psi}_M\geq 0,
    \label{eq:clausius_avg}
\end{equation}
where subscript $V$ denotes volume averages, e.g. $\bm{\sigma}_V = \smash{\frac{1}{|V|}\int_V \bm{\sigma}(\mathbf{x})\, \text{d}\mathbf{x}}$, and subscript $M$ denotes mass-weighted averages, e.g. $\smash{\dot{\psi}_M} = \smash{\frac{1}{M} \int_V \rho(\mathbf{x}) \dot{\psi}(\mathbf{x})\, \text{d}\mathbf{x}}$, with $M=\int_V \rho(\mathbf{x})\, \text{d}\mathbf{x}$ representing the total mass and $\mathbf{x}\in V$ the spatial coordinates. Note that, under the assumption of infinitesimal strains, time differentiation commutes with the averaging operators. Exploiting the formalism presented in Section~\ref{sec:theory}, inequality~\eqref{eq:clausius_avg} provides equivalent state laws for the chemical potential, energetic stress, and driving forces, namely
\begin{equation}
    \mu_V \equiv \rho_V \frac{\partial \psi_M}{\partial \rho_V}\left(\rho_V,\bm{\varepsilon}^{{e}}_V,\bm{\alpha}_V\right), \quad \bm{\sigma}^{{e}}_V \equiv \rho_V \frac{\partial \psi_M}{\partial \bm{\varepsilon}^{{e}}_V}\left(\rho_V,\bm{\varepsilon}^{{e}}_V,\bm{\alpha}_V\right), \quad \bm{\mathcal{A}}^{{e}}_V = -\bm{\mathcal{A}}^{{d}}_V \equiv \rho_V \frac{\partial \psi_M}{\partial \bm{\alpha}_V}\left(\rho_V,\bm{\varepsilon}^{{e}}_V,\bm{\alpha}_V\right).
\end{equation}
This results in equivalent expressions for the material stress and evolution equations
\begin{equation}
     \bm{\sigma}_V = \bm{\sigma}_V^e-\rho_V \mu_V\mathbf{I}, \quad \bm{z}_V = \bm{\mathbb{L}}_V \bm{\mathcal{Y}}_V,
\end{equation}
under the assumption of a negligible dissipative stress.
The volume-averaged generalised fluxes and forces, $\bm{z}_V$ and $\bm{\mathcal{Y}}_V$, collect the rates of the material-averaged irreversible strain and internal variables and the averaged generalised forces, respectively. Meanwhile, $\bm{\mathbb{L}}_V$ acts as an equivalent transport operator, whose symmetric component must be positive semi-definite to fulfil the dissipation inequality. For conciseness, the subscripts $M$ and $V$ are removed henceforth and all quantities are considered volume averages, except for the free energy which is considered as mass-weighted average. 

The macroscopic stress response extracted from DEM may exhibit fluctuations due to finite-size effects and evolving contact networks. In the following, this variability is treated as part of the observed macroscopic response (aleatoric uncertainty) and the learning task is performed on ensembles of independent realisations of the material, rather than attempting to identify or track grain-scale state variables.

\subsubsection{In silico experiments}
Granular packings are generated by first creating random clouds of spherical particles, initially with no active contacts, enclosed in a three-dimensional periodic cell. The assembly is then isotropically consolidated to prescribed stress conditions. During this preparation stage, the friction coefficient is temporarily reduced to a value $\mu_c$, with $0\leq \mu_c\leq\mu$, in order to control compaction and densification of the packing, with smaller $\mu_c$ values producing denser packings. 
\begin{figure}[ht]
\centering
\includegraphics[width=0.5\textwidth]{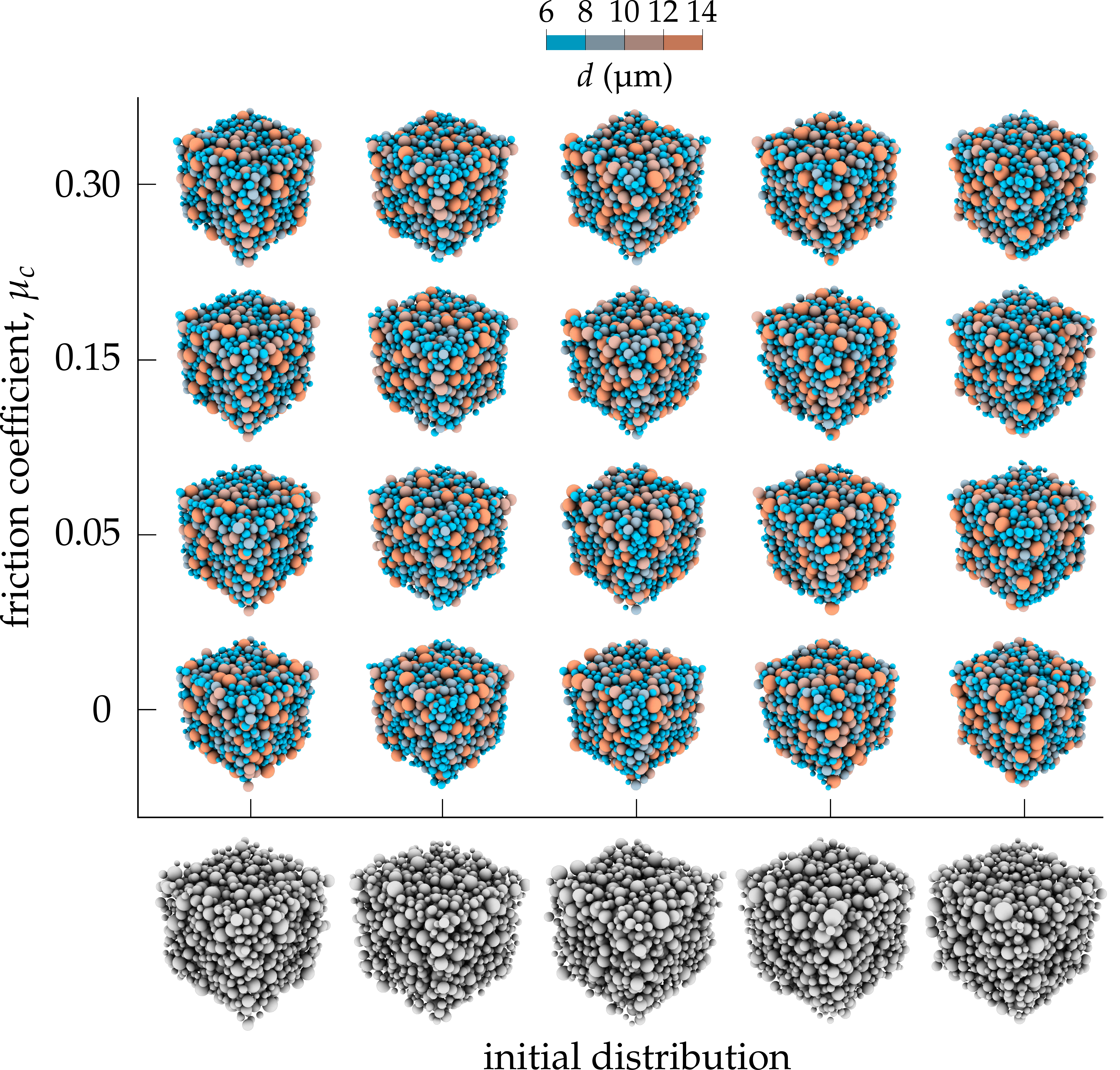}
\caption{Sample generation: (\emph{i}) creation of random particle clouds with different seeds (horizontal axis) and (\emph{ii}) compaction with different values of the friction coefficient $\mu_c$ to control the degree of densification (vertical axis). Five randomly sampled particle clouds are considered. Each cloud is subjected to isotropic consolidation (up to $p=100$ kPa) considering four different values of $\mu_c$, resulting in 20 different samples, with the same particle-size distribution.
}
\label{fig:dgen_DEM}
\end{figure}

Five independent random particle clouds are generated using different seeds. For each cloud, four different values of the preparation friction $\mu_c\in\{0,\,0.05,\,0.15,\,0.30\}$ are considered to mimic the aleatoric variability of the material. This procedure yields $20$ realisations of the same granular material, all sharing the same particle-size distribution but exhibiting different microstructures and solid fractions $\rho/\rho_s$, see Figure~\ref{fig:dgen_DEM}. 

\begin{figure}[ht]
\centering
\includegraphics[width=0.855\textwidth]{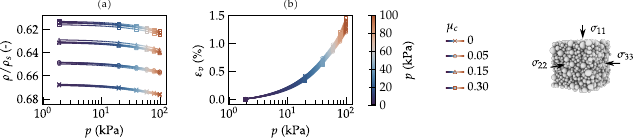}
\caption{Isotropic extension loading protocols for the training process in terms of (a) solid fraction versus isotropic pressure ($\rho/\rho_s$--$p$) and (b) volumetric strain versus isotropic pressure ($\varepsilon_v$--$p$). The measurement data include stress--strain pairs and initial density for five different microstructural realisations at $p=100$ kPa, obtained from four different values of the interparticle friction coefficient $\mu_c$.}
\label{fig:dgen_DEM_isotropic}
\end{figure}

\begin{figure}[ht]
\includegraphics[width=0.855\textwidth]{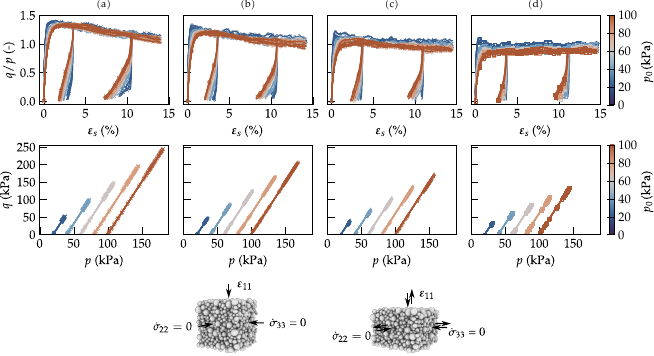}
\caption{Drained triaxial protocols for the training process in terms of (top) deviatoric stress versus strain ($q$--$\varepsilon_s$) and (bottom) deviatoric versus isotropic stress ($q$--$p$) for samples generated using a friction coefficient $\mu_c$ equal to (a) $0$, (b) $0.05$, (c) $0.15$, and (d) $0.3$. The measurement data include: (\emph{i}) monotonic triaxial compression ($\varepsilon_{11}$ prescribed, $\dot{\sigma}_{22} = \dot{\sigma}_{33} = 0$) up to $\varepsilon_s\approx 15\%$, and triaxial compression up to (\emph{ii}) $\varepsilon_s\approx 4\%$ and (\emph{iii}) $\varepsilon_s\approx 12\%$ followed by unloading up to approximately zero deviatoric stress, at different initial confining pressures $p\in(20,40,60,80,100)$ kPa.}
\label{fig:dgen_DEM_triaxial}
\end{figure}

Consolidation of the particle clouds proceeds to a homogeneous mean effective stress $p=100~\text{kPa}$ and zero deviatoric stress $q=0~\text{kPa}$, after which the contact friction is reset to its nominal value $\mu=0.5$. Each packing is then subjected to (\emph{i}) isotropic extension to generate states at different confining pressures $p \in \{20, 40, 60, 80, 100\}$ kPa, see Figure~\ref{fig:dgen_DEM_isotropic}. The resulting isotropic states, with confining pressure $0\leq p\leq 100$ kPa and solid fraction $0.6\leq \rho/\rho_s\leq 0.7$, serve as initial conditions for drained triaxial compression tests, see Figure~\ref{fig:dgen_DEM_triaxial} and \citep{masi_alert}. These tests include: (\emph{ii}) monotonic loading up to a total axial strain of approximately $15\%$, (\emph{iii}) monotonic loading to $4\%$ axial strain followed by unloading to $q=0$, and (\emph{iv})  monotonic loading to $12\%$ axial strain followed by unloading to $q=0$. The presence of non-monotonic loading paths enables the assessment of the method's ability to distinguish between loading and unloading paths. However, neither reloading nor cyclic loading is considered in the generation of the datasets used in the training process.

\subsection{Learning an effective constitutive model from macroscopic data}
\label{subsec:train_DEM}

The learning task uses only labelled volume-averaged stress--strain pairs and the initial value for the density, as in Section~\ref{subsec:training}, for both the monotonic isotropic extension tests and the (non-)monotonic triaxial tests, cf. Figures~\ref{fig:dgen_DEM_isotropic}--\ref{fig:dgen_DEM_triaxial}. 
Stress and strain are expressed through the reduced invariant representations $\bm{\sigma}_{I} = (p \quad q)^{\top}$ and $\bm{\varepsilon}_I = (\varepsilon_v \quad \varepsilon_s)^{\top}$, respectively, which are sufficient for the axisymmetric stress states covered by the triaxial protocols.

Both the free-energy network $\hat{\psi}_{\bm{\theta}}$ and the transport-operator network $\hat{\;\bm{\mathbb{L}}}_{\bm{\theta}}$ are implemented in terms of the state-space vector and generalised-force coordinates, see Section~\ref{subsec:hard_constraints}, as feed-forward neural networks with three hidden layers, 64 neurons per layer, and Softplus activation functions. The governing initial value problem is integrated using an explicit midpoint scheme, with step size $h=2.5\cdot10^{-3}$, i.e., 400 steps per protocol. 

Initial conditions are treated in a way that mirrors the data generation and loading protocols. At the reference isotropic state $p=100$ kPa, the unknown elastic strain is determined for each realisation by solving Eq.~\eqref{eq:init_cond}, with $\bm{\alpha}_0=\boldsymbol{0}$. The remaining initial states are handled by a shooting strategy, with time integration performed in a single batched solve. For the triaxial tests at lower confining pressures, $p\in\{20,40,60,80\}~\text{kPa}$, the unobserved initial elastic strain and internal variables are treated as auxiliary learnable parameters, while the initial density and macroscopic stress are prescribed from the reference data. These initial states are optimised together with the network weights by enforcing the initial stress residual in Eq.~\eqref{eq:init_cond}. Trajectories are integrated independently under their own loading conditions, but are coupled in the loss through a penalty enforcing state continuity between the connected isotropic and triaxial protocols.

\begin{figure}[h]
\includegraphics[width=0.7776\textwidth]{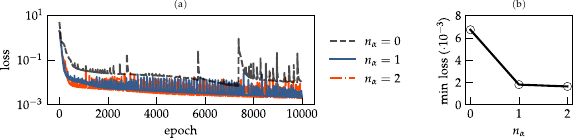}
\caption{Selection of the number of internal variables. (a) Training loss versus epochs for $n_\alpha=0,1,2$. (b) Minimum training loss as a function of $n_\alpha$.}
\label{fig:DEM_training_loss}
\end{figure}

The number of internal variables is selected according to the parsimony criterion of Section~\ref{subsubsec:internal_var}. Neural models with $n_\alpha=0,1,2$ are trained on the same dataset and compared in Figure~\ref{fig:DEM_training_loss}. Without internal variables, the model does not reproduce the observed response accurately, indicating that density and elastic strain alone are insufficient to describe the material state. A single scalar internal variable substantially improves the fit, whereas a second one brings only marginal gains. The subsequent granular-media results therefore use $n_\alpha=1$.

\subsection{Results}
All results presented hereafter are ensemble averages over five independent microstructural realisations generated as described in Section~\ref{subsec:DEM_exp}. Shaded regions indicate the corresponding 95\% confidence intervals, with light grey bands referring to the DEM simulations and dark grey bands to the predictions of the learned constitutive model. The predictions are obtained by reproducing the mixed stress- and strain-control conditions of the corresponding in silico tests.

\subsubsection{Predictions on observed loading paths}
\begin{figure}[ht!]
\includegraphics[width=0.7776\textwidth]{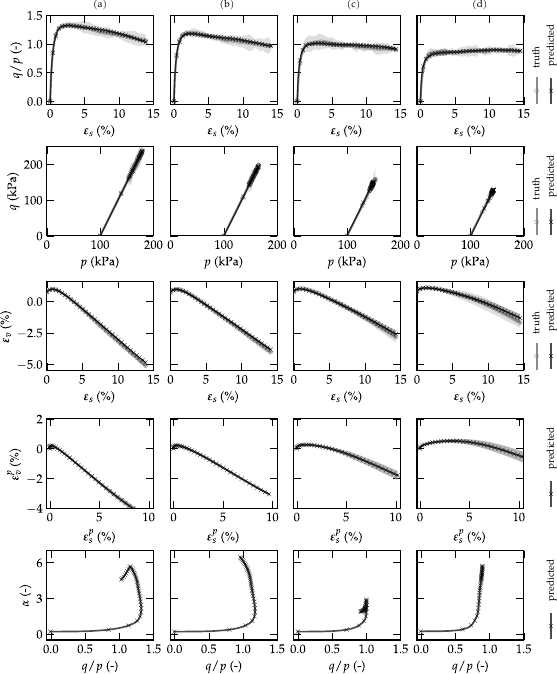}
\caption{Training predictions for monotonic drained triaxial compression tests at $p_0=100~\mathrm{kPa}$. Panels correspond to different compaction friction coefficients: 
(a) $\mu_c=0$ and $\rho_0/\rho_s=0.676$, 
(b) $\mu_c=0.05$ and $\rho_0/\rho_s=0.657$, 
(c) $\mu_c=0.15$ and $\rho_0/\rho_s=0.640$, 
(d) $\mu_c=0.3$ and $\rho_0/\rho_s=0.624$. 
From top to bottom, the rows show the stress ratio $q/p$ versus shear strain $\varepsilon_s$, the deviatoric stress $q$ versus mean pressure $p$, the volumetric strain $\varepsilon_v$ versus $\varepsilon_s$, the plastic volumetric strain $\varepsilon_v^p$ versus $\varepsilon_s^p$, and the internal state variable $\alpha$ versus $q/p$.}
\label{fig:dem_train_pred100kPa}
\end{figure}

\begin{figure}[h!]
\includegraphics[width=0.7776\textwidth]{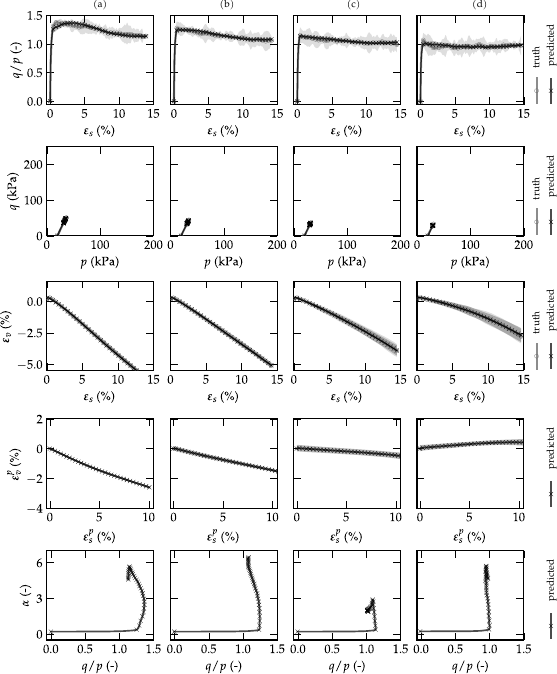}
\caption{Training predictions for monotonic drained triaxial compression tests at $p_0=20~\mathrm{kPa}$. Panels correspond to different compaction friction coefficients: 
(a) $\mu_c=0$, 
(b) $\mu_c=0.05$, 
(c) $\mu_c=0.15$, 
(d) $\mu_c=0.3$. 
From top to bottom, the rows show the stress ratio $q/p$ versus shear strain $\varepsilon_s$, the deviatoric stress $q$ versus mean pressure $p$, the volumetric strain $\varepsilon_v$ versus $\varepsilon_s$, the plastic volumetric strain $\varepsilon_v^p$ versus $\varepsilon_s^p$, and the internal state variable $\alpha$ versus $q/p$.}
\label{fig:dem_train_pred20kPa}
\end{figure}

Figures~\ref{fig:dem_train_pred100kPa} and \ref{fig:dem_train_pred20kPa} show the response under monotonic drained triaxial compression tests, included in the training dataset, starting from $p_0=100\,\mathrm{kPa}$ and $p_0=20\,\mathrm{kPa}$, respectively. The different columns correspond to simulated granular packings obtained with different compaction friction coefficients $\mu_c$, and hence with different initial densities and microstructural states. The first three rows of Figures~\ref{fig:dem_train_pred100kPa} and \ref{fig:dem_train_pred20kPa} compare the observable macroscopic quantities from the numerical experiments with the learned-model predictions. The neural constitutive model reproduces the main density-dependent trends across the observed densities. Dense packings, obtained for smaller $\mu_c$, exhibit a higher peak stress ratio $q/p$ followed by softening, whereas looser packings show a lower peak and a smoother approach to the $q/p$ plateau. The agreement in the $(q,p)$ plane further indicates that the learned state law retrieves the coupling between deviatoric and mean stress during drained triaxial loading.
The volumetric response follows from the enforcement of the drained boundary conditions. The model captures the transition from initial compaction to subsequent dilation and its dependence on density. The agreement is particularly clear for dense and intermediate packings. For looser states, the predicted mean response remains close to the reference trend, although a small drift appears at larger strains.

The last two rows report latent quantities inferred by the model, namely the plastic-strain coordinates $(\varepsilon_v^p,\varepsilon_s^p)$ and the scalar internal variable $(\alpha)$. These quantities are not supervised during training and are not uniquely identifiable from macroscopic data alone. Nevertheless, their evolution is consistent across protocols. The inferred plastic strains follow the loading direction and reflect the contractive or dilative trends of the macroscopic response, while the internal variable remains nearly constant during the early stages of loading and changes more rapidly as the stress ratio approaches its peak and subsequent plateau. Together with the stress predictions, this suggests that the learned state representation captures the main path-dependent features of the DEM responses.

\subsubsection{Generalisation to unseen loading paths}
\begin{figure}[ht!]
\includegraphics[width=0.7776\textwidth]{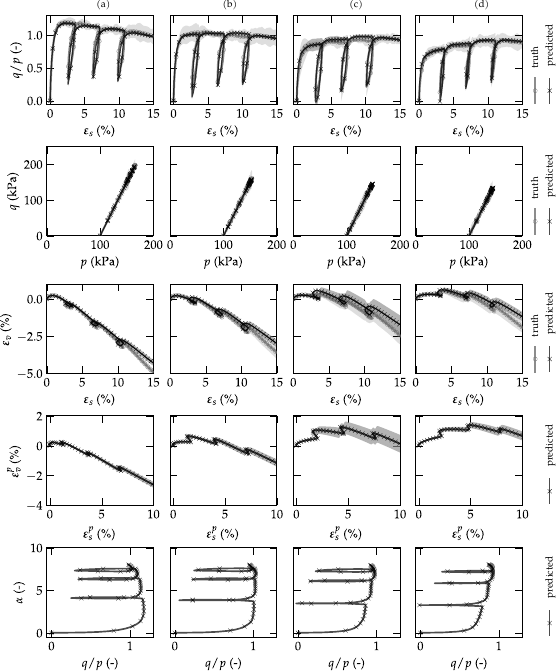}
\caption{Inference predictions for unobserved drained triaxial compression tests with three loading--unloading sequences at $p_0=100~\mathrm{kPa}$. Panels correspond to different interparticle preparation friction coefficients:
(a) $\mu_c=0.05$,
(b) $\mu_c=0.15$,
(c) $\mu_c=0.3$,
(d) $\mu_c=0.5$, i.e., $\rho_0/\rho_s=0.616$.
From top to bottom, the rows show the stress ratio $q/p$ versus shear strain $\varepsilon_s$, the deviatoric stress $q$ versus mean pressure $p$, the volumetric response $\varepsilon_v$ versus $\varepsilon_s$, the plastic volumetric strain $\varepsilon_v^p$ versus $\varepsilon_s^p$, and the internal variable $\alpha$ versus $q/p$.}
\label{fig:dem_pred_cyclic3}
\end{figure}

The generalisation capabilities of the learned model are evaluated considering unseen loading paths and protocols, with increasing difficulty of extrapolation. Figure~\ref{fig:dem_pred_cyclic3} shows drained triaxial compression tests with three loading--unloading sequences at $p_0=100~\mathrm{kPa}$ and different compaction friction coefficients $\mu_c$. The model predicts the unloading--reloading branches with good accuracy, despite having never observed reloading during training. The predicted $q/p$--$\varepsilon_s$ branches reproduce the material response during unloading and reloading, including the changes in stiffness induced by strain reversals, and the $(q,p)$ paths remain close to the numerical experiments throughout the loading sequence. The total volumetric response is also captured well, with the model predicting the accumulated volumetric dilation and the shifts induced by multiple reversals, with only minor local discrepancies with respect to the DEM simulations.

The predicted plastic strain $(\varepsilon_v^p,\varepsilon_s^p)$ paths develop distinct branches during loading and unloading, while $\alpha$ follows separated trajectories as the stress ratio is reversed and reloaded. This behaviour suggests that $\alpha$ provides the model with a scalar memory variable needed to distinguish loading from unloading and reloading.

It is worth noting that Figure~\ref{fig:dem_pred_cyclic3}~(d) considers a looser packing ($\mu_c=0.5$) than those used in training, for which $\mu_c\leq0.3$. Nevertheless, the model still captures the main stress-ratio and volumetric trends, demonstrating that the learned dependence on density and state variables provides a meaningful extrapolation beyond the observed compaction conditions.\\

\begin{figure}[ht!]
\includegraphics[width=0.7776\textwidth]{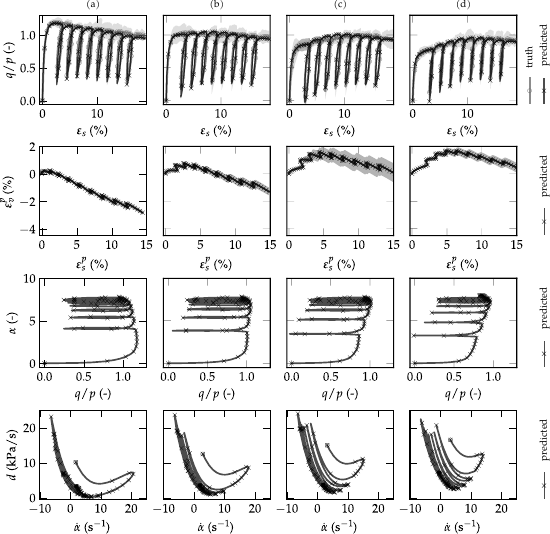}
\caption{Inference predictions for unobserved drained triaxial compression tests with eight loading--unloading sequences at $p_0=100~\mathrm{kPa}$. Panels correspond to different interparticle preparation friction coefficients:
(a) $\mu_c=0.05$,
(b) $\mu_c=0.15$,
(c) $\mu_c=0.3$,
(d) $\mu_c=0.5$.
From top to bottom, the rows show the stress ratio $q/p$ versus shear strain $\varepsilon_s$, the plastic volumetric strain $\varepsilon_v^p$ versus $\varepsilon_s^p$, the internal variable $\alpha$ versus $q/p$, and the dissipation rate $d$ versus the internal-variable rate $\dot{\alpha}$. Open circles mark the initial value of the $d$--$\dot{\alpha}$ trajectory, while filled circles mark its final value.}
\label{fig:dem_pred_cyclic8}
\end{figure}

Figure~\ref{fig:dem_pred_cyclic8} considers the same initial configurations as Figure~\ref{fig:dem_pred_cyclic3}, but increases the number of unloading--reloading sequences from three to eight, with smaller strain amplitudes. The stress-ratio response remains stable over these repeated cycles, and the predicted branches follow the main path-dependent trends of the in silico experiments. Moreover, the model does not show visible spurious drift over the tested horizon, despite not having been trained on reloading, cyclic paths, or long time horizons. The latent variables in Figure~\ref{fig:dem_pred_cyclic8} reveal how the model organises the repeated unloading--reloading sequence. The internal variable $\alpha$ increases during the first loading stages and then evolves within a bounded range, approaching an asymptotic state rather than growing indefinitely. This suggests that $\alpha$ encodes the progressive rearrangement of the granular state until a nearly recurrent cyclic response is reached. Consistently, the trajectories in the $d$--$\dot{\alpha}$ plane remain non-negative and form cycle-by-cycle loops that drift during the transient stages before approaching a more stable recurrent pattern.\\

\begin{figure}[h!]
\includegraphics[width=0.603\textwidth]{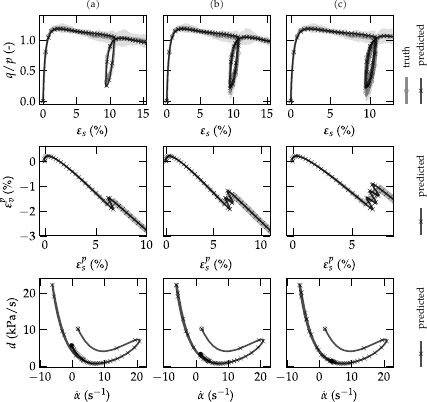}
\caption{Inference predictions for unobserved cyclic drained triaxial compression tests at $p_0=100~\mathrm{kPa}$ and $\mu_c=0.05$. Panels correspond to protocols with an increasing number of repeated cycles: (a) one unloading--reloading cycle, (b) two cycles, and (c) three cycles. From top to bottom, the rows show the stress ratio $q/p$ versus shear strain $\varepsilon_s$, the plastic volumetric strain $\varepsilon_v^p$ versus $\varepsilon_s^p$, and the dissipation rate $d$ versus the internal-variable rate $\dot{\alpha}$. Open circles mark the initial value of the $d$--$\dot{\alpha}$ trajectory, while filled circles mark its final value.}
\label{fig:dem_pred_hysteresis}
\end{figure}

Figure~\ref{fig:dem_pred_hysteresis} presents unloading--reloading cycles in drained triaxial compression. The three columns correspond to protocols with an increasing number of repeated cycles: (a) one unloading--reloading cycle, (b) two cycles, and (c) three cycles. These protocols produce clear hysteretic loops in the $q/p$--$\varepsilon_s$ plane, which are well captured by the model.

The neural model recovers the initial loading branch, the unloading and reloading paths, and the progressive formation of closed loops as the number of cycles increases, with only a slight underestimation of the gradual reduction of $q/p$, before reversal. The same behaviour is reflected in the inferred plastic strain trajectories. 

The dissipation-rate plots provide a thermodynamic view of the same cyclic evolution. The predicted trajectories in the $d$--$\dot{\alpha}$ plane organise into closed limit cycles. This result is particularly significant because neither plastic strains nor dissipation rates are directly supervised during training, yet they carry qualitative information about the nonlinear hysteretic behaviour of the material.\\

\begin{figure}[ht!]
\includegraphics[width=0.873\textwidth]{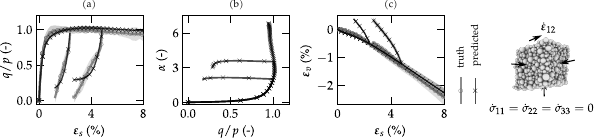}
\caption{Inference predictions for an unobserved simple-shear test with unloading at $p_0=100~\mathrm{kPa}$ and $\mu_c=0.15$. From left to right: (a) stress ratio $q/p$ versus shear strain $\varepsilon_s$, (b) internal variable $\alpha$ versus $q/p$, and (c) volumetric strain $\varepsilon_v$ versus shear strain $\varepsilon_s$.}
\label{fig:dem_pred_shear}
\end{figure}

Finally, Figure~\ref{fig:dem_pred_shear} considers a loading protocol absent from the training dataset: simple shear with unloading, in which $\varepsilon_{12}$ is prescribed while the normal stress components are kept fixed, i.e. $\dot{\sigma}_{11}=\dot{\sigma}_{22}=\dot{\sigma}_{33}=0$. This mixed-control condition is enforced through the algebraic stress constraints of the constitutive update.

Despite being trained only on isotropic and triaxial paths, the model accurately reproduces the stress-ratio response during shear loading and, to a certain extent, during unloading, with some discrepancies as the unloading progresses. 

The model also captures the overall volumetric trend in the $(\varepsilon_v,\varepsilon_s)$ response, but overestimates the contraction after reversal. This limitation is expected because simple shear probes a loading mode fundamentally outside the triaxial protocols used for training and suggests that a wider class of protocols is required to improve extrapolation under non-triaxial conditions.

Taken together, Figures~\ref{fig:dem_pred_cyclic3}--\ref{fig:dem_pred_shear} show that the identified constitutive equations generalise beyond the observed monotonic and unloading paths, recovering reloading, hysteretic cycles, and different loading protocols. Particularly interesting is the evolution of the automatically identified internal variable $\alpha$, which provides a compact latent coordinate through which the model represents material path dependence, while remaining microstructure-agnostic.

\section{Conclusions}
\label{sec:conclusions}
This study develops a new framework to learn and discover constitutive models for complex, inelastic materials grounded in non-equilibrium thermodynamics \citep{de2013non} and, in particular, in the framework of generalised transport equations \citep{gurtin1996generalized,van2008internal}. The constitutive closure is built on two coupled neural parametrisations: a free energy \citep{masi2021thermodynamics, masi2022multiscale} and a state- and force-dependent transport operator that governs the evolution of an automatically identified material state space. Here, the first law of thermodynamics is enforced through an energetic parametrisation of the state laws, and the second law is satisfied exactly by restricting the symmetric part of the transport operator to be positive semidefinite \citep{van2008internal}, while retaining possible skew-symmetric couplings to represent reversible interactions and mixed reversible--irreversible effects. The framework additionally hard-wires frame indifference through invariant representations and uses convexity in selected invariant coordinates as a stability-motivated energetic constraint, while automatically identifying a stress-free reference equilibrium state. Unlike soft-constraint approaches that may drift into thermodynamic violations when trained on limited datasets (cf. \citep{masi2024neural}), and unlike many hard-constrained formulations that rely on restrictive templates (e.g. dissipation pseudo-potentials \citep{halphen1975materiaux,mcbride2018dissipation,de2013non}), the current approach defines an admissible yet flexible hypothesis class able to represent inelastic materials, including non-associative and path-dependent behaviours. The framework also removes the need to access state variables, such as elastic (or plastic) strain and internal variables, that are not directly measurable from conventional laboratory experiments without additional modelling assumptions.

The performance of the framework was validated on synthetic benchmarks generated from analytical material models of increasing complexity, spanning elasto- and hypo-plasticity, rate effects, hardening, and non-associative plasticity. Across these cases, the automatically identified models reproduce the observed stress histories with high accuracy while simultaneously delivering internally consistent thermodynamic variables (free energy, dissipation rate, and state evolution) despite these quantities not being observed at training time. The benchmarks also illustrate that the method can identify interpretable internal variables when they are necessary to represent the data (e.g. a hardening-like variable emerging without supervision), and can generalise to qualitatively different and more demanding loading paths under mixed control. The framework was further applied to granular materials using discrete-element simulations as controlled analogues of experiments, with prescribed particle-scale geometry and contact behaviour. The training data were restricted to macroscopic stress--strain histories and initial density, so that grain-scale descriptors remained unobserved. The identified constitutive model reproduces monotonic and non-monotonic responses and extrapolates to selected unseen cyclic loading paths and different protocols, including hysteresis absent from the training data.

While the presented framework relies on a Cauchy continuum description, restricted to isothermal processes, the general non-equilibrium thermodynamic setting it builds upon can be naturally extended to generalised continua \citep{germain1973method,maugin1990internal,ireman2004using,papenfuss2006thermodynamical}, i.e., nonlocal and higher-order formulations. In this context, the dual internal variable theory \citep{van2008internal,berezovski2018internal}, which proposes a unifying description of internal variables and internal degrees of freedom, offers a compelling standpoint to enrich the state space without relying on ad hoc closures (cf. \citep{ottinger2005beyond}). 

The proposed framework is general in its thermodynamic structure and can be applied to a wide range of materials with complex microstructural effects. The results demonstrate that accurate macroscopic representations can be identified in a microstructure-agnostic setting. Whenever microstructural observations are accessible, the same approach can be coupled to automatically discovered descriptors of the evolving microstructure to promote identifiability of the discovered internal state variables (see \citep{masi2022multiscale,masi2023evolution}).
 
Application of the present methodology to real experimental observations is a natural next step. In that setting, the hard-constrained structure is expected to act both as a robustness mechanism under noisy and sparse measurements and as an explicit diagnostic of thermodynamic admissibility under extrapolation.

\subsubsection*{Supplementary material}
The implementation of the present learning framework is available at {\small \url{github.com/filippo-masi/constitutive-transport-learning}}.

\subsubsection*{Acknowledgements}
I would like to thank Prof. Ioannis Stefanou for fruitful discussions, guidance, and collaboration within the framework of the CoQuake project (European Research Council, Grant Agreement ID 757848), which helped shape the scientific questions addressed in this work and motivated the proposed framework. I would also like to thank Prof. Itai Einav for insightful discussions on the theoretical foundations of the Onsager--Casimir reciprocity relations, and Dr. Vincent Acary for valuable exchanges on the implementation of the proposed methodology.\\
This work was supported by the Multidisciplinary Institute in Artificial Intelligence (MIAI) Cluster and the \emph{Agence Nationale de la Recherche} through the France 2030 program (Grant agreement ANR-23-IACL-0006) within the chair AIM: Artificial Intelligence and Mechanics for scale bridging in complex materials.

\subsubsection*{Declaration of competing interests}
The author declares no competing interests.

\subsubsection*{Declaration of generative AI and AI-assisted technologies}
During the preparation of this manuscript, the author used OpenAI's large language model, ChatGPT, to assist with text refinement and grammar checking. After using this tool, the author reviewed and edited the content as needed and takes full responsibility for the content of the present manuscript.

\appendix
\renewcommand{\theequation}{\thesection.\arabic{equation}}
\renewcommand{\thefigure}{\thesection.\arabic{figure}}
\numberwithin{equation}{section}

\section{Classification based on the total strain}
\label{appendix:strain}
One may prefer not to assume the existence of elastic strain, to any extent, when setting a general classification for the material state space. This section presents the derivation of Section~\ref{sec:theory} with a state space specified as ${\bm{s}}=(\rho, \bm{\varepsilon}, \bm{\alpha})$, where the total strain is used as a state variable.
Consequently, the free energy and the energetic stress are specified as
\begin{equation}
\psi \equiv \psi\left(\rho,\bm{\varepsilon},\bm{\alpha}\right), \quad {\bm{\sigma}}^{e} \equiv \rho \frac{\partial \psi}{\partial \bm{\varepsilon}}\left(\rho,\bm{\varepsilon},\bm{\alpha}\right),
\label{eq:energy_app}
\end{equation}
while the chemical potential and energetic forces associated with the internal variables are defined in Eq.~\eqref{eq:conj}. Following the same procedure as in Section~\ref{sec:theory}, the dissipation inequality reads
\begin{equation}
 d =  \bm{\mathcal{A}}^{{d}} \bm{\cdot} \dot{\bm{\alpha}} + \bm{\sigma}^{{d}} \!:\!\dot{\bm{\varepsilon}} = {\bm{\mathcal{Y}}}\bm{\cdot} {\bm{z}} \geq 0,\qquad  {\bm{\mathcal{Y}}} \equiv \begin{pmatrix}
\bm{\mathcal{A}}^{{d}} & \bm{\sigma}^{{d}}
\end{pmatrix}^{\top} ,\quad {\bm{z}} \equiv 
\begin{pmatrix}
\dot{\bm{\alpha}} & \dot{\bm{\varepsilon}}
\end{pmatrix}^{\top},
\label{eq:dissipation_app}
\end{equation}
where $\bm{\sigma}^{{d}} \equiv \bm{\sigma} -  \big[{\bm{\sigma}}^{{e}}-\rho{\mu}\mathbf{I}\big]$ and $\bm{\mathcal{A}}^{d} \equiv - \rho \partial \psi/\partial \bm{\alpha}$.
A general solution to the dissipation inequality is thus given by ${\bm{z}} = {\bm{\mathbb{L}}}{\bm{\mathcal{Y}}}$, i.e.,
\begin{equation}
    \begin{pmatrix}
    \dot{\bm{\alpha}}\\
    \dot{\bm{\varepsilon}}
    \end{pmatrix}=
    \begin{pmatrix}
    {\mathbf{L}}^{{11}} & {\mathbf{L}}^{{12}} \\
    {\mathbf{L}}^{{21}} & {\mathbf{L}}^{{22}}\\
    \end{pmatrix}
    \begin{pmatrix}
    \bm{\mathcal{A}}^{{d}} \\
    \bm{\sigma}^{{d}}
    \end{pmatrix}
\end{equation}
with $\operatorname{Sym}\big({\bm{\mathbb{L}}}\big)\succeq 0$.

\section{Hard versus soft enforcement of the dissipation inequality}
\label{app:hard_soft_dissipation}
The impact of enforcing the dissipation inequality by construction (hard constraint) versus enforcing it through a penalty term (soft constraint) is assessed on the one-dimensional elasto-plastic benchmark with isotropic hardening (Section~\ref{subsubsec:EPH}). For completeness, two penalty-based alternatives for the evolution law are considered, namely
\begin{itemize}
    \item soft-direct: a direct evolution network (cf. \citep{masi2024neural}), i.e., ${\bm{z}}_I=\bm{g}_{\bm{\theta}}(\bm{s}_I,\bm{\mathcal Y}_I)$ and
    \item soft-transport: a transport relation ${\bm{z}}_I=\!\hat{\;\bm{\mathbb{L}}}_{\bm{\theta}}(\bm{s}_I,\bm{\mathcal Y}_I)\,\bm{\mathcal Y}_I$ where the symmetric part of $\hat{\;\bm{\mathbb{L}}}_{\bm{\theta}}$ is not constrained to be positive semidefinite.
\end{itemize}

In both cases, the objective in Eq.~\eqref{eq:loss} is augmented with a penalty term for negative dissipation rate,
\begin{equation}
    \bm{\theta}^\star = \arg\min_{\bm{\theta}} \frac{1}{n_{\text P}}\sum_{\ell=1}^{n_{\text P}}\left[\frac{1}{n_{\text T}}\sum_{k=1}^{n_{\text T}}\left\|{\bm{\sigma}}_I(\bm{s}_{I,k}^{\psup}; \bm{\theta})-\bar{\bm{\sigma}}^{\psup}_{I,k}\right\|_2^{2} + \left\|\bm{\mathcal{R}}(\bm{s}_{I,0}^{\psup}, \bar{\bm{\sigma}}_{I,0}^{\psup}; \bm{\theta})\right\|_2^{2}+\frac{1}{n_{\text T}}
\sum_{k=1}^{n_{\text T}} \left[\max\!\left(0,-\tilde{d}(\bm{s}_{I,k}^{\psup};\bm{\theta})\right)\right]^2\right],
\end{equation}
where $d(\bm{s}_I;\bm{\theta})$ denotes the dissipation rate computed as the inner product between the predicted thermodynamic forces and fluxes, and $\tilde d=d/d_0$ is its normalised value, with $d_0$ a characteristic dissipation scale computed from the training data. As in Eq.~\eqref{eq:loss}, the stress errors and initial-condition residuals are evaluated in normalised stress-invariant coordinates.

All comparisons use the same free-energy architecture, state-law parametrisations, training protocols, optimiser settings, numerical integration scheme, initialisation, and random seeds, as well as the same number of trainable parameters for the evolution model, see Section~\ref{subsubsec:EPH}. Training is performed for 4000 epochs for all three variants.

\begin{figure}[th]
\centering
\includegraphics[width=0.63\textwidth]{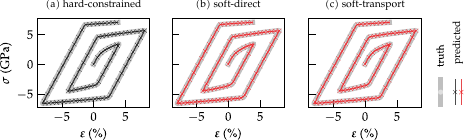}
\caption{Stress--strain response for one training loading protocol: (a) hard-constrained formulation, (b--c) soft-constrained variants. All formulations achieve comparable fits on the training data.}
\label{fig:abla_1}
\end{figure}

Figure~\ref{fig:abla_1} compares the three formulations (hard-constrained, soft-direct, and soft-transport) for one of the training loading protocols. All formulations achieve comparable stress--strain fits on the training data, indicating that the observed stress response alone does not strongly discriminate between enforcement strategies for the dissipation inequality or between the chosen evolution parametrisations.

\begin{figure}[th]
\centering
\includegraphics[width=0.6975\textwidth]{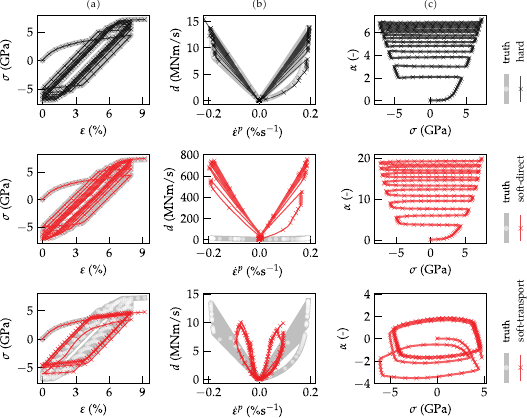}
\caption{Predictions on the unseen cyclic loading protocol of Fig.~\ref{fig:EPH_inf} for the hard-constrained (top) and soft-constrained implementations, soft-direct (middle) and soft-transport (bottom). From left to right: (a) stress versus strain ($\sigma$--$\varepsilon$), (b) dissipation rate versus plastic strain rate ($d$--$\dot{\varepsilon}^p$), (c) automatically identified internal variable $\alpha$. Panel (b) uses different vertical axis limits across rows to accommodate the larger dissipation magnitudes of the soft-direct variant.}
\label{fig:abla_2}
\end{figure}

Generalisation is then assessed on the unseen cyclic loading protocol presented in Figure~\ref{fig:EPH_inf}. Figure~\ref{fig:abla_2} compares predictions with the ground truth. Although the soft-constraint variants exhibit non-negative dissipation rates for this protocol, they yield degraded extrapolation performance in terms of (\emph{i}) the stress hysteresis, (\emph{ii}) the dissipation rate $d$, and (\emph{iii}) the evolution of the learned internal variable. This indicates that penalising violations of $d\ge 0$ during training can be insufficient to constrain the learned evolution operator outside the training distribution: multiple evolution laws can fit the training data while remaining weakly constrained and may fail to generalise.  By contrast, the hard-constrained formulation remains admissible by construction and provides more reliable extrapolation.

{
\setlength{\bibsep}{2pt plus 0.3ex}
\bibliography{biblio}
}

\end{document}